%% file: main1.tex
\documentclass[12pt]{article}

\usepackage[numbers, sort&compress]{natbib}
\usepackage{fullpage}
\usepackage{epsfig}
\usepackage{latexsym,amsmath,amssymb,amsthm}
\usepackage{subfig}
\newcommand{\ceil}[1]{\left\lceil {#1} \right\rceil}

\def\maps11{\stackrel {1-1}{\longmapsto}}

\begin{document}

\title{Scheduling strategies and throughput optimization for the Uplink
for IEEE 802.11ax and IEEE 802.11ac based networks}

\author{%
Oran Sharon
\thanks{Corresponding author: oran@netanya.ac.il, Tel: 972-4-9831406,
Fax: 972-4-9930525} \\
Department of Computer Science \\
Netanya Academic College \\
1 University St. \\
Netanya, 42365 Israel
\and
Yaron Alpert\\
Intel\\
13 Zarchin St.\\
Ra'anana, 43662, Israel\\
Yaron.alpert@intel.com
}


\date{}

\maketitle

\begin{abstract} 
The new IEEE 802.11 standard, IEEE 802.11ax, 
has the challenging
goal of serving more Uplink (UL) traffic and users as compared
with his predecessor IEEE 802.11ac,
enabling consistent and reliable 
streams of data (average throughput) per station.
In this paper we explore several new
IEEE 802.11ax UL scheduling mechanisms and
compare between the maximum throughputs 
of unidirectional
UDP Multi Users (MU) triadic. The evaluation is
conducted based 
on Multiple-Input-Multiple-Output (MIMO) 
and Orthogonal Frequency Division Multiple Acceess (OFDMA)
transmission multiplexing format 
in IEEE 802.11ax vs. the CSMA/CA MAC in IEEE 802.11ac in 
the Single User (SU) and MU modes for 1, 4, 
8, 16, 32 and 64 stations scenario in reliable and unreliable
channels. The comparison is conducted
as a function of the Modulation and Coding
Schemes (MCS) in use. In IEEE 802.11ax 
we consider two new flavors of
acknowledgment operation settings, 
where the maximum acknowledgment
windows are 64 or 256 respectively. 
In SU scenario the throughputs of IEEE 802.11ax
are larger than those of IEEE 802.11ac 
by 64$\%$ and 85$\%$ in reliable
and unreliable channels respectively. 
In MU-MIMO scenario the throughputs of
IEEE 802.11ax are larger than those of
IEEE  802.11ac by 263$\%$ and
270$\%$ in reliable and unreliable channels 
respectively.
Also, as the number
of stations increases, the advantage 
of IEEE 802.11ax in terms of the access
delay also increases.
\end{abstract}
\bigskip


\renewcommand{\baselinestretch}{1.3}
\small\normalsize


\input intro

\input newax
\input model

\input thrana
\input thrres

\input summary

\clearpage


\bibliographystyle{abbrv}
\bibliography{main}


\end{document}

%% file: intro.tex
\section{Introduction}

\subsection{Background}

The latest IEEE 802.11 Standard (WiFi)~\cite{IEEEBase1}, 
created and maintained by 
the IEEE LAN/MAN Standards Committee (IEEE 802.11),
is currently the most effective solution within the range of Wireless Local
Area Networks (WLAN). Since its first release 
in 1997 the standard provides the basis 
for Wireless network products using 
the WiFi brand, and has since been 
improved upon in many ways. 
One of the main goals of these improvements
is to increase the throughput 
achieved by users and to improve
the standard's Quality-of-Service (QoS) capabilities. 
To fulfill the promise of increasing 
IEEE 802.11 performance and QoS capabilities,
a new amendment, IEEE 802.11ax ( also
known as High Efficiency (HE) ) was recently 
introduced~\cite{IEEEax}. IEEE 802.11ax is 
considered to be the sixth generation 
of a WLAN in the IEEE 802.11 set 
of types of WLANs and it is 
a successor to IEEE 802.11ac~\cite{IEEEac,PS}.
The scope of the IEEE 802.11ax amendment is to
define modifications for both the 802.11 PHY and MAC
layers that enable at least four-fold improvement
in the average throughput per station in densely
deployed networks~\cite{KKL, AVA, DCC, B}.
Currently IEEE 802.11ax project is 
in a very early stage of development and  
is due to be publicly released in 2019 .

\subsection{Research question}

In order to achieve its goals, one of 
the main challenges of IEEE 802.11ax is to enable simultaneous
transmissions by several stations and to enable
Quality-of-Service. 
In this paper we
assume that the AP
is communicating in a regular fashion with a fix set of stations.
We explore some of the UL IEEE 802.11ax new 
mechanisms given that the AP knows with
which stations it communicates and we compare 
between the unidirectional 
UDP throughputs of IEEE 802.11ax and IEEE 802.11ac 
in Single User (SU) and Multi User (MU) modes
for 1, 4, 8, 16, 32 and 64 
stations scenarios in reliable and
unreliable channels. 
This is one of the aspects to compare between new
amendments of the IEEE 802.11 standard~\cite{KCC}.
The SU scenario 
implements sequential transmissions
in which a single wireless station sends
and receives data at every cycle one at 
a time, once it or the AP has gained access 
to the medium. The MU scenarios allow
for simultaneous transmission and reception to and 
from  multiple stations both in the 
Downlink (DL) and UL directions. 
UL MU refers to simultaneous transmissions, i.e. at the same
time, from several stations to the AP over the UL.
The existing IEEE 802.11ac standard
does not enable UL MU while IEEE 802.11ax enables
up to 74 stations to transmit simultaneously over the UL.

The MU transmissions over the UL are done
by MIMO and Orthogonal Frequency Divisionn Multiple
Access (OFDMA).
The IEEE 802.11ax standard
expends MIMO transmissions multiplexing format 
and specifies new ways of multiplexing additional users 
using OFDMA. 
The new IEEE 802.11ax OFDMA is backward compatible 
and enables scheduling different users in different
sub-carriers of the same channel. 
In the IEEE 802.11ac the total channel 
bandwidth (20 MHz, 40 MHz, 80 MHz etc. ) 
contains multiple OFDM sub-carriers. 
However, in IEEE 802.11ax OFDMA, different 
subsets of sub-carriers in the channel 
bandwidth can be used by different frame 
transmissions at the same time. Sub-carriers 
can be allocated for transmissions 
in Resource Units (RU)  as small as 2 MHz.

Given the above new structure of OFDMA in IEEE 802.11ax, 
the main contributions of this paper are as follows: First
we suggest several scheduling strategies by which a given
number of stations can transmit over the UL. Second, we evaluate
the throughput and access delay performance of the
different scheduling strategies given the different PHY rates
of the RUs in the various scheduling strategies, and
the different number of RUs in use, which influences the
PHY preamble's length.




\subsection{Previous works}

Most of the research papers on IEEE 802.11ax
so far deal with these challenges and examine different
access methods to enable efficient multi user access
to random sets of stations.
For example, 
in~\cite{QLYY} the authors deal with the introduction
of OFDMA
into IEEE 802.11ax to enable multi user access.
They introduce an OFDMA based multiple access
protocol, denoted Orthogonal MAC for 802.11ax (OMAX),
to solve synchronization
problems and to reduce the overhead associated with
using OFDMA.
In~\cite{LLYQYZY} the authors suggest an access protocol
over the UL of an IEEE 802.11ax WLAN based on Multi User
Multiple-Input-Multiple-Output (MU-MIMO) and OFDMA PHY.
In~\cite{LDC} the authors suggest a centralized medium
access protocol for the UL of IEEE 802.11ax in order to efficiently
use the transmission resources.
In this protocol stations transmit
requests for frequency sub-carriers, denoted
Resource Units (RU) to the AP over the UL. The AP
allocates RUs to the stations which use them later
for data transmissions over the UL.
In~\cite{KBPSL} a new method to use OFDMA over the UL
is suggested, where MAC Protocol Data Units (MPDU) from the stations are
of different lenghs.
In~\cite{JS, RFBBO, RBFB, HYSG} a new version 
of the Carrier Sense Multiple Access with Collision Avoidance
(CSMA/CA) protocol, denoted Enhanced CSMA/CA (CSMA/ECA) is
suggested, which is suitable for IEEE 802.11ax . A deterministic
backoff is used after a successful transmission, and the backoff
stage is not reset after service. The backoff stage is reset
only when a station does not have any more MPDUs to transmit.
CSMA/ECA enables a more efficient use of the channel
and enhanced fairness.
In~\cite{KLL} the authors assume
a network with legacy and IEEE 802.11ax stations and examine
fairness issues between the two sets of the stations. 

\indent
The rest of the paper is organized as follows: 
In Section 2 we describe
the new mechanisms of IEEE 802.11ax 
relevant to this paper. In Section 3 we 
describe the transmission scenario 
with which we compare between IEEE 802.11ax 
and IEEE 802.11ac in the SU and MU modes.
We assume the reader is familiar 
with the basics of the PHY and MAC layers
of IEEE 802.11 described in previous papers, e.g.~\cite{SA}. 
In Section 4 we analytically compute 
the IEEE 802.11ax and IEEE 802.11ac 
throughputs. 
In Section 5 we present the throughput 
of the various protocols and compare between
them. Section 6 summarizes 
the paper. In the rest of the paper 
we denote IEEE 802.11ac and IEEE 802.11ax 
by 11ac and 11ax respectively.

%% file: newax.tex
\section{The new features in IEEE 802.11ax}

IEEE 802.11ax focuses on implementing 
mechanisms to serve more
users simultaneously, enabling consistent and 
reliable streams of data ( average throughput
per user ) in the presence of many other users. 
Therefore there are several 
new mechanisms in 11ax compared to 11ac both in the PHY and
MAC layers. At the PHY layer, 
11ax enables larger OFDM FFT sizes, 4X larger,
therefore every OFDM symbol is 
extended from $3.2 \mu s$ in 11ac to $12.8 \mu s$ in 11ax. 
By narrower subcarrier spacing (4X closer)
the protocol efficiency is increased because the
same Guard Interval (GI) is used both in 11ax and 11ac .

Additionally, to increase the average 
throughput per user in high-density scenarios,
11ax expends the 11ac Modulation Coding Schemes 
(MCSs) and adds MCS10 (1024 QAM ) and MCS 11 (024 QAM 5/6), 
applicable for transmission with bandwidth larger than 20 MHz.

In this paper we focus on UL scheduling methods
that enable to optimize
the IEEE 802.11 two-level aggregation
schemes working point, first introduced 
in IEEE 802.11n~\cite{IEEEBase1,PS}, 
in which several  MPDUs can be aggregated 
to be transmitted in a  single PHY Service 
Data Unit (PSDU). Such aggregated PSDU 
is denoted Aggregate MAC Protocol Data
Unit (A-MPDU) frame. In two-level aggregation 
every MPDU can contain several 
MAC Service Data Units (MSDU). 
MPDUs are separated by an MPDU 
Delimiter field of 4 bytes and 
each MPDU contains MAC Header
and Frame Control Sequence (FCS) fields.
MSDUs within an MPDU are separated 
by a SubHeader field of 14 bytes. 
Every MSDU is rounded to an 
integral multiple of 4 bytes 
together with the SubHeader field. 
Every MPDU is also rounded to 
an integral multiple of 4 bytes.

\begin{figure}
\vskip 9cm
\includegraphics{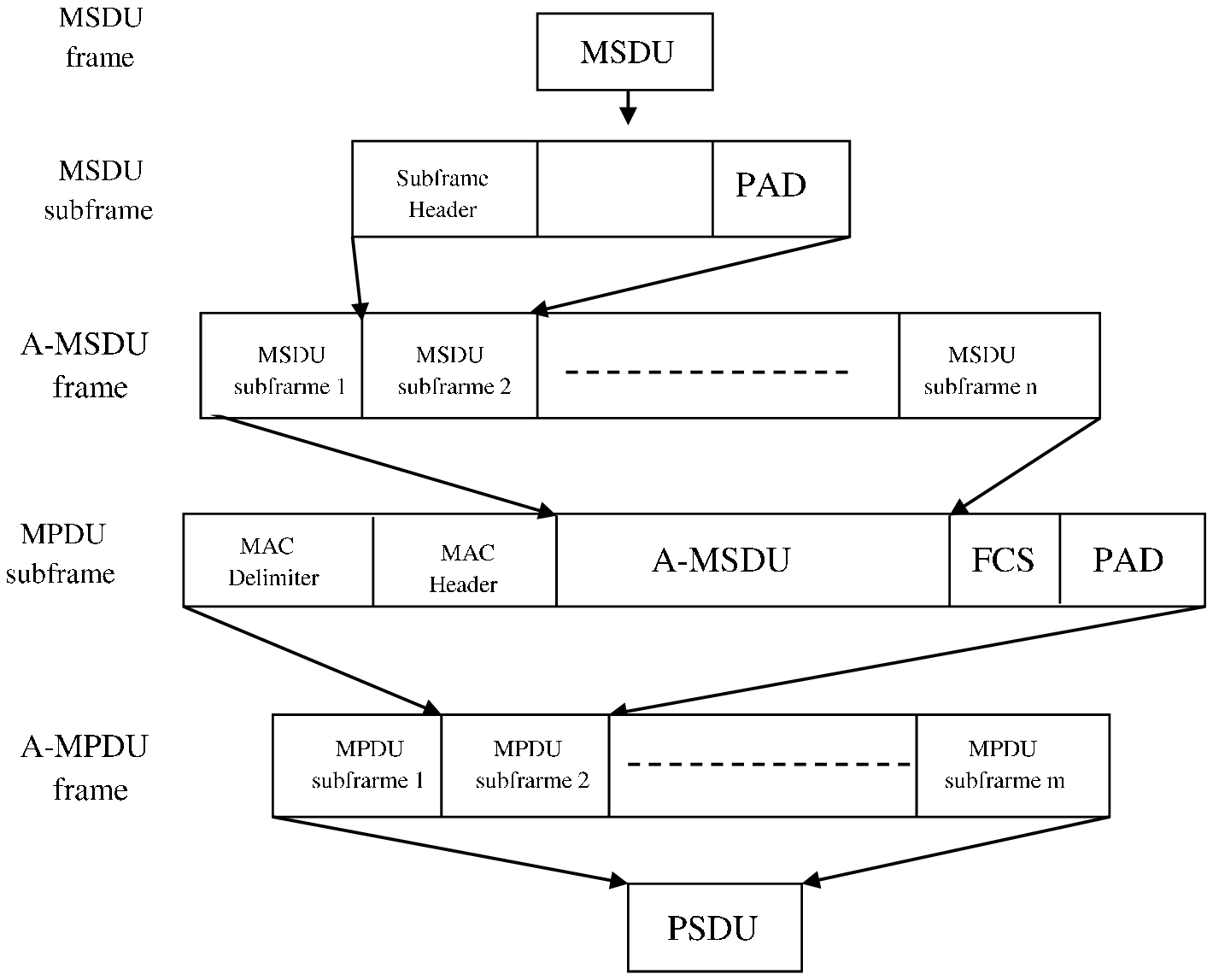}
\caption{The generation of an A-MPDU frame
in two-level aggregation.}
\label{fig:twole}
\end{figure}

In 11ax and 11ac the size of 
an MPDU is limited to 11454 bytes. In 11ac an A-MPDU
is limited to 1,048,575 bytes and 
this limit is extended to 4,194,304 bytes 
in 11ax. In both 11ac and 11ax 
the transmission time of the PPDU 
(PSDU and its preamble) is limited to 
$5.484ms$ ($5484 \mu s$)
due to L-SIG (one of the legacy 
preamble's fields) duration limit~\cite{IEEEBase1}.
The A-MPDU frame structure in two-level aggregation
is shown in Figure~\ref{fig:twole}.

IEEE 802.11ax also enables the extension of the acknowledgment 
mechanism by using a 256 maximum 
acknowledgment window vs. maximum 
window of 64 in 11ac. In this paper 
we also assume that all MPDUs 
transmitted in an A-MPDU frame are 
from the same Traffic Stream (TS). 
In this case up to 256 MPDUs are
allowed in an A-MPDU frame of 11ax, 
while in 11ac up to only 64 MPDUs
are allowed. 

The acknowledgments are transmitted by
special control frames, Block Ack (BAck) and
Multi Station BAck, to be specified later.

Finally, in 11ac it 
is not possible to transmit simultaneously 
over the UL and only SU is supported.
In 11ax this is possible using MU and
up to 74 stations can transmit simultaneously.

%% file: model.tex
\section{Model}

\subsection{Transmission patterns}

\begin{figure}
\vskip 19cm
\includegraphics{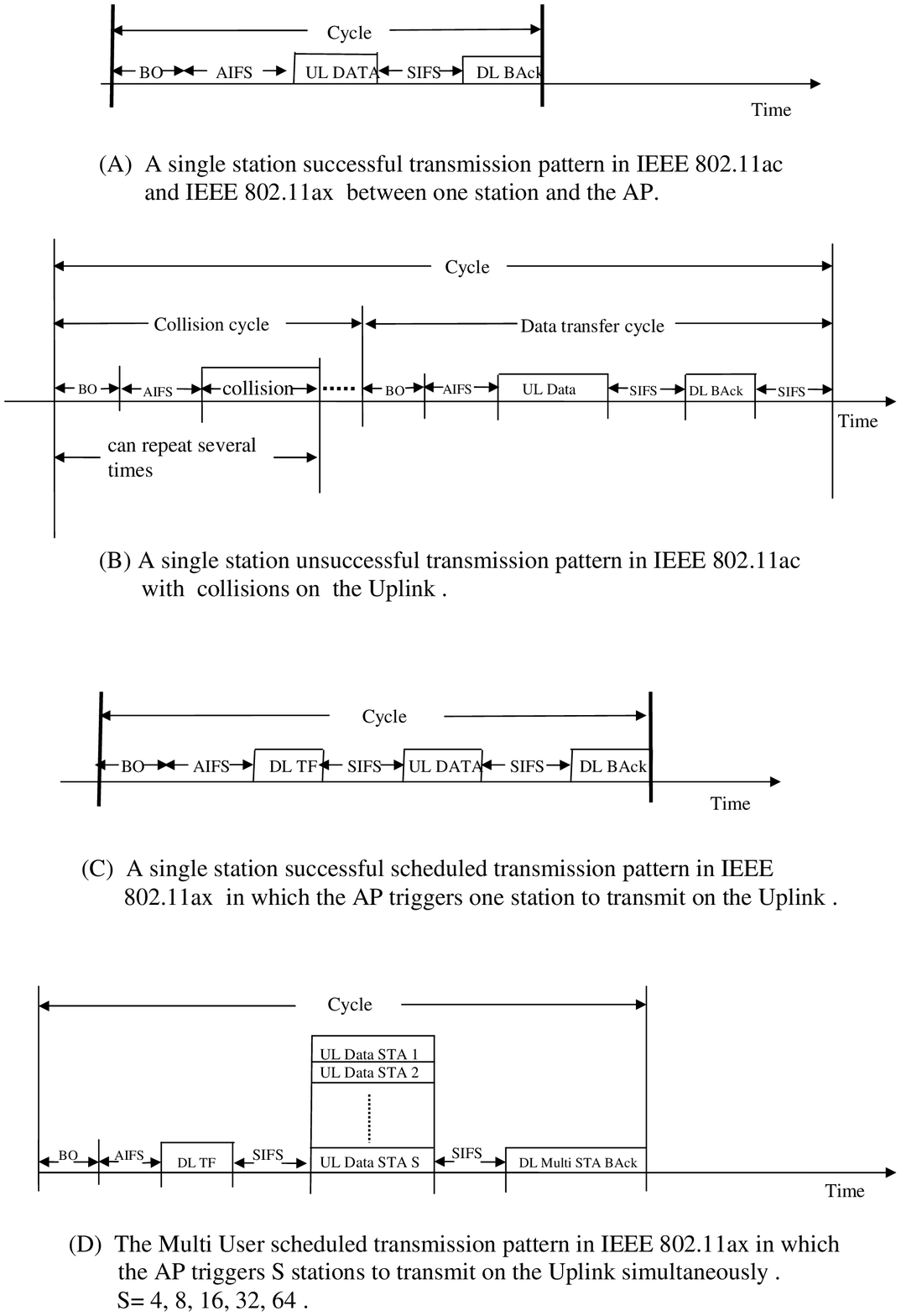}
\caption{Transmissions from stations to the AP in Single User and Multi User modes in IEEE 802.11ac and in IEEE 802.11ax .}
\label{fig:traffic}
\end{figure}

One of the main goals of 11ax is to enable
larger throughputs in the network when
several stations are transmitting simultaneously
over the UL to the AP. 
In 11ax it is possible to use the MU
transmission mode over the UL and up to 74
stations can transmit simultaneously 
to the AP.
On the other hand 11ac
does not support the MU transmission mode
on the UL, thus 
when several
stations need to transmit they must access the channel
one by one, using the CSMA/CA MAC with possible collisions.
In this paper we compare
between the throughputs received in 11ac and 11ax
over the UL
when $S$ stations, $S=1, 4, 8, 16, 32$ and
64 stations transmit 
to the AP, which in turn replies with a common
MAC acknowledgment frame, BAck or Multi Station BAck to be
specified later.
In both 11ac and 11ax when
only one station is transmitting in the system, this
is done by the Single User (SU) mode
of transmissions.
The station transmits data frames to the AP
and receives common broadcast
BAck frames in return. In this mode
the advantage of the 11ax over 11ac is due to its more
efficient PHY layer and its new MCSs. The UL traffic
pattern in this case is shown in Figure~\ref{fig:traffic}(A)
for both 11ac and 11ax.

When several stations are transmitting over the UL
in 11ac, the air access selection is
done by using the CSMA/CA MAC
only, which involves collisions between stations, as shown
in Figure~\ref{fig:traffic}(B). 
In 11ax such transmissions can be done
by either SU (similar to 11ac method) or new UL MU-MIMO modes 
under control of the AP. The transmission pattern in 11ax
using SU is shown in
Figure~\ref{fig:traffic}(C) which repeats itself $S$ times
when $S$ stations transmit.
The AP allocates resources (RU) and
solicits the stations to transmit
by a spacial control frame,
Trigger Frame (TF).
Another 11ax MU transmission alternative is to use 
a combination of UL MU-MIMO and OFDMA
in which several stations transmit simultaneously
in the same transmission opportunity over the UL.
In Figure~\ref{fig:traffic}(D) we show this possibility for 11ax
where the AP is communicating with $S>1$ stations.

In the case of UL MU
the AP allocates the UL Resource Units (RU), i.e.
Frequency/Spatial Streams (SS) for transmission
of the stations
by the before-mentioned TF control frame
which is transmitted over the DL
to the stations.
In the TF, the AP allocates UL RUs and defines the UL transmission
format per station. Following the UL transmission the AP
acknowledges reception of the data frames by transmitting
a common new control frame, Multi Station BAck, to the
stations. In this common frame the AP transmits Ack
information per station, as a response to the last UL transmission.

\begin{figure}
\vskip 15cm
\includegraphics{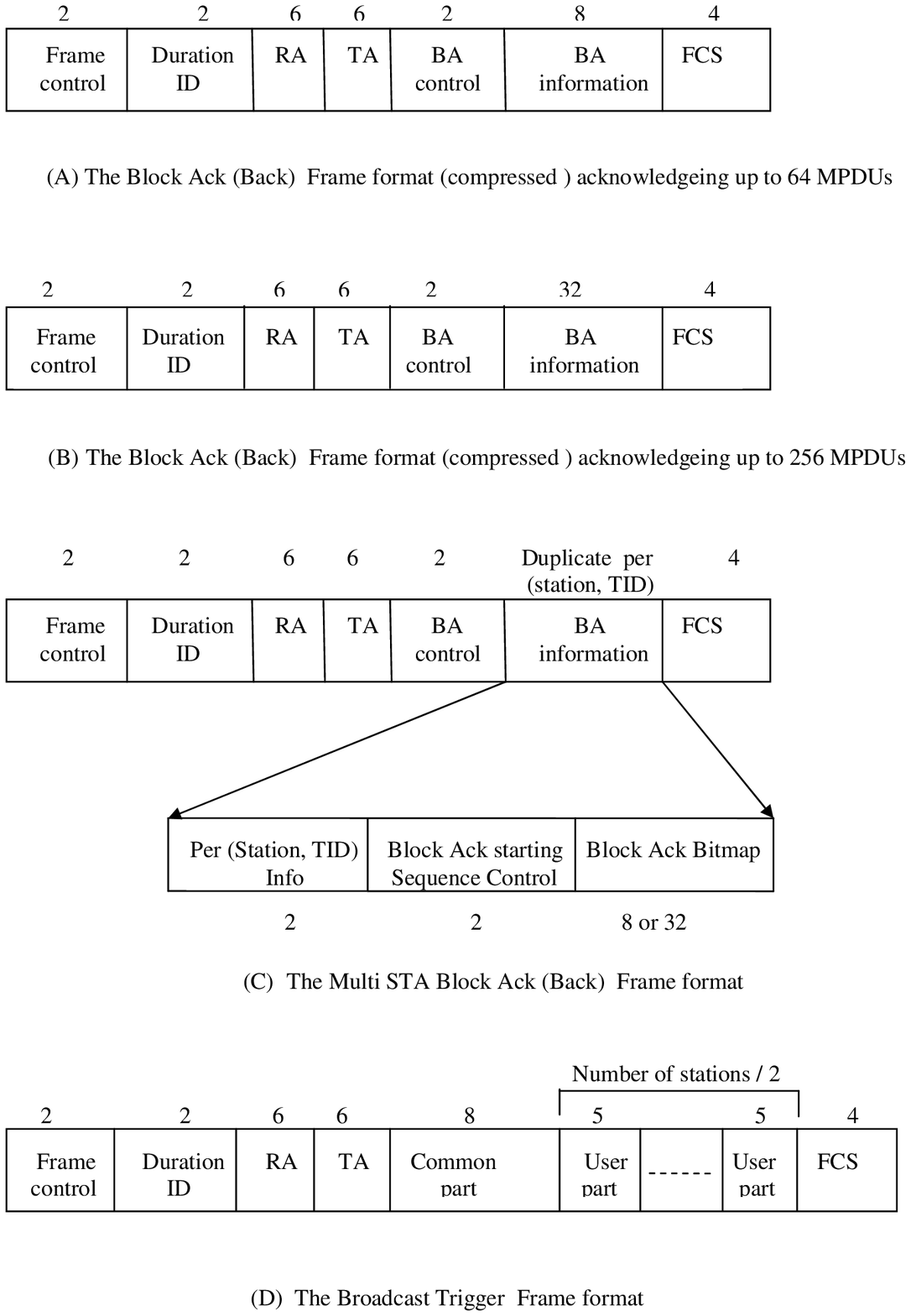}
\caption{The Block Ack,  Multi Station Block Ack and Trigger Frame frames' formats.} 
\label{fig:frameformat}
\end{figure}

The AP uses the legacy transmission mode when transmitting
the BAck, the TF and the Multi Station BAck 
frames over the DL, both in 11ac and 11ax.
The formats
of the BAck, Multi Station BAck and TF frames are shown in
Figure~\ref{fig:frameformat}(A), (B), (C) and (D) respectively.
The difference between the BAck frames is Figures~\ref{fig:frameformat}(A)
and (B) is that the former acknowledges up to 64 MPDUs
while the latter acknowledges up to 256 MPDUs. This format
is applicable in 11ax only.

\newpage

\subsection{UL Transmissions' service scheduling strategies}

In 11ax 
there are several UL scheduling and
non-scheduling service strategies for the stations
to transmit data to the AP,
and we compare between
them. Recall that in 11ac the only possible service strategy
is to use the CSMA/CA MAC over the UL, as shown
in Figure~\ref{fig:traffic}(B).

We now specify the UL service scheduling strategies in 11ax for every
number $S$ of stations, $S=1, 4, 8, 16, 32, 64$.
By $SU_{AX}$ we refer to the traffic pattern in 
Figure~\ref{fig:traffic}(A).
By $x \cdot SU_{AX}(1)$
we denote a transmission by $n$ stations
in 11ax
using the transmission pattern in Figure~\ref{fig:traffic}(C)
$x$ times in sequence; every transmission is by a different
station. 
By $m \cdot MU_{AX}(n)$ we denote transmissions
by $m \cdot n$ stations using the traffic pattern 
of Figure~\ref{fig:traffic}(D) $m$ times in sequence; each transmission
is by a different group of $n$ stations. In this paper
$n=4, 8, 16, 32$ and 64.

\noindent
The UL service scheduling strategies are as follows:

\begin{itemize}

\item
$S=1$:

\noindent
 $1 \cdot SU_{AX}$ .

\item
$S=4$:

\noindent
 $4 \cdot SU_{AX}(1)$, $1 \cdot MU_{AX}(4)$.

\item
$S=8$:

\noindent
 $8 \cdot SU_{AX}(1)$, $2 \cdot MU_{AX}(4)$, $1 \cdot MU_{AX}(8)$ .

\item
$S=16$:

\noindent
 $16 \cdot SU_{AX}(1)$, $4 \cdot MU_{AX}(4)$, $2 \cdot MU_{AX}(8)$, $1 \cdot MU_{AX}(16)$.

\item
$S=32$:

\noindent
 $32 \cdot SU_{AX}(1)$, $8 \cdot MU_{AX}(4)$, $4 \cdot MU_{AX}(8)$, $2 \cdot MU_{AX}(16)$, $1 \cdot MU_{AX}(32)$.

\item
$S=64$:

\noindent
 $64 \cdot SU_{AX}(1)$, $16 \cdot MU_{AX}(4)$, $8 \cdot MU_{AX}(8)$, $4 \cdot MU_{AX}(16)$, $2 \cdot MU_{AX}(32)$, $1 \cdot MU_{AX}(64)$.

\end{itemize}

\subsection{Channel assignment}

We assume
the 5GHz band, a 160MHz channel and that the AP 
and the stations have 4 antennas each.
In 11ac every station transmits using 4 SSs.
This is because 11ac supports UL SU only
and a single station can transmit in all 4 SS 
if needed.
In 11ax a station transmits in SU mode,
Figure~\ref{fig:traffic}(A) and~\ref{fig:traffic}(C),
by using 4 SSs and
in MU mode by using 1 SS.
Recall that in both 11ac and 11ax, 
and in both SU and MU modes in 11ax,
the AP transmits over the DL by using the legacy mode.
The DL PHY rate is usually set to the minimum between the UL
Data rate and the largest possible PHY rate in the set 
of the basic rates that
is smaller or equal to the UL Data rate.
The minimal basic PHY rate is 6Mbps and in the case
of UL PHY rates smaller than 6Mbps
the DL PHY rate is never less than 6Mbps. This can happen in case
of 64 stations (see Table 2).

When using the MU mode in 11ax, the 160MHz channel
is divided in the UL into $\frac{S}{4}$ channels of
$\frac{160 \cdot 4}{S}$ MHz each, $S=4, 8, 16, 32, 64$.
In every such channel
4 stations transmit to the AP, each
using 1 SS. For example, for $S=64$
there are 16 channels of 10MHz each; in each of them
4 stations transmit to the AP.
When $S=4$ only MU is used. For $S>4$ MU-MIMO+OFDMA is used.

\subsection{PPDU formats}

\begin{figure}
\vskip 13cm
\includegraphics{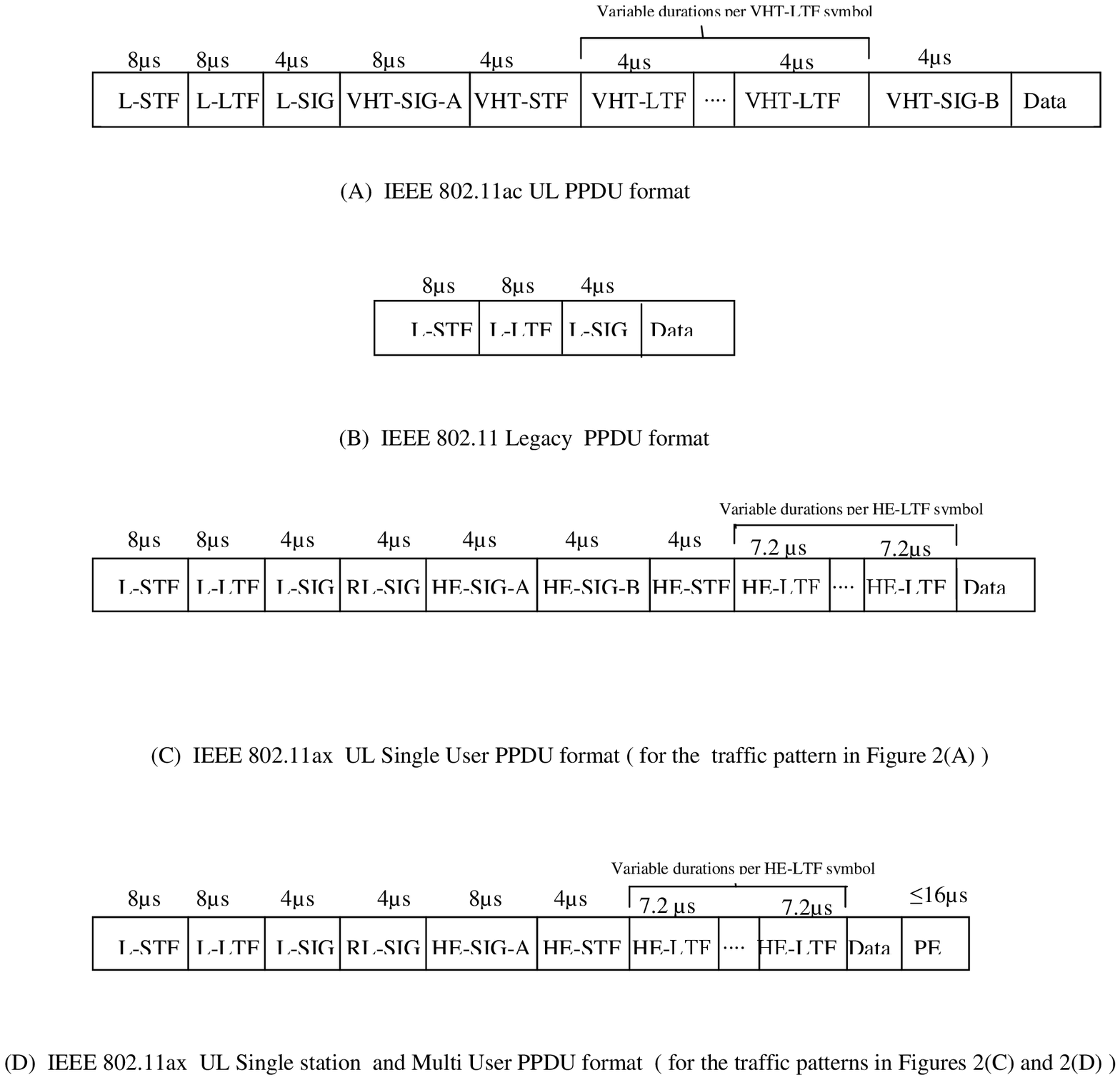}
\caption{The PPDU formats in the SU and MU modes.}
\label{fig:formatPPDU}
\end{figure}

In Figure~\ref{fig:formatPPDU} we show the various
PPDUs' formats in use in the
various transmission patterns of Figure~\ref{fig:traffic}.

In Figure~\ref{fig:formatPPDU}(A) 
we show the PPDU format
used over the UL in the traffic pattern of 11ac , 
Figures~\ref{fig:traffic}(A) and~\ref{fig:traffic}(B).
In this PPDU format there are
the VHT-LTF fields, the number of which equals
the number of SSs in use (4 in our case), and each is $4 \mu s$ .

In Figure~\ref{fig:formatPPDU} (B) we
show the legacy preamble, used in both 11ac and 11ax over the
DL. 

In Figure~\ref{fig:formatPPDU} (C) we show the PPDU format used in
11ax UL SU mode, Figure~\ref{fig:traffic}(A).

In Figure~\ref{fig:formatPPDU} (D)
we show the PPDU format
used over the UL  when a single 
station transmits in 11ax, Figure~\ref{fig:traffic}(C),
and UL MU transmission patterns of 11ax , 
Figure~\ref{fig:traffic}(D).

In the 11ax PPDU format there are the HE-LTF fields,
the number of which equals the number of SSs
in use, 4 in our case.
In this paper we assume that each such field is
composed of 2X LTF and therefore of duration
$7.2 \mu s$~\cite{IEEEax}.

Notice also that the PSDU frame in 11ax contains
a Packet Extension (PE) field.
This field is mainly used in MU mode
and we assume that it is $0 \mu s$ in 
SU and the longest possible in MU,
$16 \mu s$.

\subsection{Parameters' values}

In Table~\ref{tab:phyratesSU} we show the
PHY rates and the length of the preambles that
are used in 11ac and 11ax in SU mode and
in the various MCSs. The values are taken from~\cite{IEEEax}.

\begin{table}
\caption{\label{tab:phyratesSU}{The PHY rates and the preambles in the UL and DL
of IEEE 802.11ac and IEEE 802.11ax in Single User mode. A 160 MHz channel, 4 Spatial Streams in the UL. DL Ack conducted at the basic rate set.}} 
\vspace{3 mm}
\tiny
\center
\begin{tabular}{|r|c|c|c|c|c|c|c|c|c|}  \hline
     & \multicolumn{2}{|c|}1 &  \multicolumn{2}{|c|}2  &  &
       \multicolumn{2}{|c|}3 & \multicolumn{2}{|c|}4 \\ \cline{2-10}
     & \multicolumn{2}{c|}{SU UL data}&  \multicolumn{2}{c|}{SU UL data}&  &
       \multicolumn{2}{c|}{DL BAck transmission} & \multicolumn{2}{c|}{DL BAck transmission} \\
     & \multicolumn{2}{c|}{transmission rate in 11ax}  &  \multicolumn{2}{c|}{transmission rate in 11ac}&  &
       \multicolumn{2}{c|}{rate for 11ax} & \multicolumn{2}{c|}{rate for 11ac}  \\ \hline
     &  PHY Rate & Preamble & 
       PHY Rate & Preamble & & PHY Rate & Preamble & PHY Rate & Preamble \\
MCS  &  (Mbps)   & ($\mu s$) 
     &  (Mbps)    & ($\mu s$)  & & (Mbps)   & ($\mu s$) & (Mbps) & ($\mu s$) \\
     &  GI$=0.8 \mu s$   &
       & GI$=0.8 \mu s$   &   & &  & & &  \\ \hline
     & \multicolumn{2}{c|}{1 station IEEE 802.11 ax} & \multicolumn{2}{c|}{ 1 station IEEE 802.11 ac} & & \multicolumn{2}{c|}{} & \multicolumn{2}{c|}{} \\ \hline
 0   &  288.2  & 60.8  &   234.0 & 52.0 && 48.0 & 20.0 & 48.0 & 20.0 \\ 
 1   &  576.5  & 60.8  &   468.0 & 52.0 && 48.0 & 20.0 & 48.0 & 20.0 \\ 
 2   &  864.7  & 60.8  &   702.5 & 52.0 && 48.0 & 20.0 & 48.0 & 20.0 \\ 
 3   & 1152.9  & 60.8  &   936.0 & 52.0 && 48.0 & 20.0 & 48.0 & 20.0 \\ 
 4   & 1729.4  & 60.8  &  1404.0 & 52.0 && 48.0 & 20.0 & 48.0 & 20.0 \\ 
 5   & 2305.9  & 60.8  &  1872.0 & 52.0 && 48.0 & 20.0 & 48.0 & 20.0 \\ 
 6   & 2594.1  & 60.8  &  2106.0 & 52.0 && 48.0 & 20.0 & 48.0 & 20.0 \\ 
 7   & 2882.4  & 60.8  &  2340.0 & 52.0 && 48.0 & 20.0 & 48.0 & 20.0 \\ 
8    & 3458.8  & 60.8  &  2808.0 & 52.0 && 48.0 & 20.0 & 48.0 & 20.0 \\ 
 9   & 3848.1  & 60.8  &  3120.0 & 52.0 && 48.0 & 20.0 & 48.0 & 20.0 \\ 
10   & 4323.5  & 60.8  &  N/A  & N/A && 48.0 & 20.0  & N/A & N/A \\ 
11   & 4803.9  & 60.8  &  N/A  & N/A && 48.0 & 20.0  & N/A & N/A \\ \hline 
\end{tabular}  
\end{table}

In Table~\ref{tab:phyrates} we show the PHY rates and the preambles
used in 11ax in MU mode, in the various MCSs and
in all cases of the number of stations $S$, i.e. 
$S= 4, 8, 16, 32$ and 64. We also include again 
the PHY rates of 11ac in SU mode which are also used 
when $S>1$ stations are transmitting over the UL.

\begin{table}
\caption{\label{tab:phyrates}{The PHY rates and the preambles in the UL MU of IEEE 802.11ax and the SU UL of IEEE 802.11ac. A 160 MHz channel, 4 Spatial Streams in the UL. DL Ack and TF transmission is conducted at the basic rate set. }}
\vspace{3 mm}
\tiny
\center
\begin{tabular}{|r|c|c|c|c|c|c|c|c|c|}  \hline
     & \multicolumn{2}{|c|}1 &  \multicolumn{2}{|c|}2  &  &
       \multicolumn{2}{|c|}3 & \multicolumn{2}{|c|}4 \\ \cline{2-10}
     & \multicolumn{2}{c|}{MU UL data}&  \multicolumn{2}{c|}{SU UL data}&  & \multicolumn{2}{c|}{DL TF/Multi Station BAck} & \multicolumn{2}{c|}{DL BAck} \\ & \multicolumn{2}{c|}{transmission rate in 11ax}  &  \multicolumn{2}{c|}{transmission rate in 11ac}&  & \multicolumn{2}{c|}{transmission rate for 11ax}  & \multicolumn{2}{c|}{transmission rate for 11ac} \\ \hline
     &  PHY Rate & Preamble & 
       PHY Rate & Preamble & & PHY Rate & Preamble & PHY Rate & Preamble \\
MCS  &  (Mbps)   & ($\mu s$) 
     &  (Mbps)    & ($\mu s$)  & & (Mbps)   & ($\mu s$) & (Mbps) & ($\mu s$) \\
     &  GI$=1.6 \mu s$   &
       & GI$=0.8 \mu s$   &   & & GI$=0.9 \mu s$ & & GI$=0.8 \mu$ &  \\ \hline
     & \multicolumn{2}{c|}{4 stations IEEE 802.11 ax} &  &\multicolumn{6}{c|}{} \\ \hline
 0   &  68.1   & 64.8  &  234.0  & 52.0 && 48.0 & 20.0 & 48.0 & 20.0 \\ 
 1   &  136.1  & 64.8  &  468.0  & 52.0 && 48.0 & 20.0 & 48.0 & 20.0 \\ 
 2   &  204.2  & 64.8  &  702.0  & 52.0 && 48.0 & 20.0 & 48.0 & 20.0 \\ 
 3   &  272.2  & 64.8  &  936.0  & 52.0 && 48.0 & 20.0 & 48.0 & 20.0 \\ 
 4   &  408.3  & 64.8  & 1404.0  & 52.0 && 48.0 & 20.0 & 48.0 & 20.0 \\ 
 5   &  544.4  & 64.8  & 1872.0  & 52.0 && 48.0 & 20.0 & 48.0 & 20.0 \\ 
 6   &  612.5  & 64.8  & 2106.0  & 52.0 && 48.0 & 20.0 & 48.0 & 20.0 \\ 
 7   &  680.6  & 64.8  & 2340.0  & 52.0 && 48.0 & 20.0 & 48.0 & 20.0 \\ 
8    &  816.7  & 64.8  & 2808.0  & 52.0 && 48.0 & 20.0 & 48.0 & 20.0 \\ 
 9   &  907.4  & 64.8  & 3120.0  & 52.0 && 48.0 & 20.0 & 48.0 & 20.0 \\ 
10   & 1020.8  & 64.8  &  N/A    &  N/A && 48.0 & 20.0 & N/A & N/A \\ 
11   & 1134.2  & 64.8  &  N/A    &  N/A && 48.0 & 20.0 & N/A & N/A  \\ \hline 
     & \multicolumn{2}{c|}{8 stations IEEE 802.11 ax} &  &\multicolumn{6}{c|}{} \\ \hline
 0   &  34.0 & 64.8  &  234.0  & 52.0 && 36.0 & 20.0 & 48.0 & 20.0 \\ 
 1   &  68.1 & 64.8  &  468.0  & 52.0 && 48.0 & 20.0 & 48.0 & 20.0 \\ 
 2   & 102.1 & 64.8  &  702.0  & 52.0 && 48.0 & 20.0 & 48.0 & 20.0 \\ 
 3   & 136.1 & 64.8  &  936.0  & 52.0 && 48.0 & 20.0 & 48.0 & 20.0 \\ 
 4   & 204.2 & 64.8  & 1404.0  & 52.0 && 48.0 & 20.0 & 48.0 & 20.0 \\ 
 5   & 272.2 & 64.8  & 1872.0  & 52.0 && 48.0 & 20.0 & 48.0 & 20.0 \\ 
 6   & 306.3 & 64.8  & 2106.0  & 52.0 && 48.0 & 20.0 & 48.0 & 20.0 \\ 
 7   & 340.3 & 64.8  & 2340.0  & 52.0 && 48.0 & 20.0 & 48.0 & 20.0 \\ 
8    & 408.3 & 64.8  & 2808.0  & 52.0 && 48.0 & 20.0 & 48.0 & 20.0 \\ 
 9   & 453.7 & 64.8  & 3120.0  & 52.0 && 48.0 & 20.0 & 48.0 & 20.0 \\ 
10   & 510.4 & 64.8  &  N/A    &  N/A && 48.0 & 20.0 &  N/A & N/A  \\
11   & 567.1 & 64.8  &  N/A    &  N/A && 48.0 & 20.0 &  N/A & N/A  \\ \hline 
     & \multicolumn{2}{c|}{16 stations IEEE 802.11 ax} &  &\multicolumn{6}{c|}{} \\ \hline
 0   &  16.3 & 64.8 &  234.0 & 52.0 && 12.0 & 20.0 & 48.0 & 20.0 \\ 
 2   &  48.8 & 64.8 &  702.0 & 52.0 && 24.0 & 20.0 & 48.0 & 20.0\\ 
 3   &  65.0 & 64.8 &  936.0 & 52.0 && 48.0 & 20.0 & 48.0 & 20.0\\ 
 4   &  97.5 & 64.8 & 1404.0 & 52.0 && 48.0 & 20.0 & 48.0 & 20.0\\ 
 5   & 130.0 & 64.8 & 1872.0 & 52.0 && 48.0 & 20.0 & 48.0 & 20.0\\ 
 6   & 146.3 & 64.8 & 2106.0 & 52.0 && 48.0 & 20.0 & 48.0 & 20.0\\ 
 7   & 162.5 & 64.8 & 2340.0 & 52.0 && 48.0 & 20.0 & 48.0 & 20.0\\ 
8    & 195.0 & 64.8 & 2808.0 & 52.0 && 48.0 & 20.0 & 48.0 & 20.0\\ 
 9   & 216.7 & 64.8 & 3120.0 & 52.0 && 48.0 & 20.0 & 48.0 & 20.0\\ 
10   & 243.8 & 64.8 &   N/A  &  N/A && 48.0 & 20.0 &  N/A & N/A \\ 
11   & 270.8 & 64.8 &   N/A  &  N/A && 48.0 & 20.0 &  N/A & N/A \\ \hline 
\end{tabular} 
\end{table}

\addtocounter{table}{-1}

\begin{table}
\caption{\label{tab:phyrates1}{(cont.)}}
\vspace{3 mm}
\tiny
\center
\begin{tabular}{|r|c|c|c|c|c|c|c|c|c|}  \hline
     & \multicolumn{2}{|c|}1 &  \multicolumn{2}{|c|}2  &  &
       \multicolumn{2}{|c|}3 & \multicolumn{2}{|c|}4 \\ \cline{2-10}
     & \multicolumn{2}{c|}{MU UL data}&  \multicolumn{2}{c|}{SU UL data}&  &
       \multicolumn{2}{c|}{DL TF/BAck} & \multicolumn{2}{c|}{DL BAck} \\
     & \multicolumn{2}{c|}{transmission rate in 11ax}  &  \multicolumn{2}{c|}{transmission rate in 11ac}&  & \multicolumn{2}{c|}{transmission rate for 11ax}  & \multicolumn{2}{c|}{transmission rate for 11ac} \\ \hline
     &  PHY Rate & Preamble & PHY Rate & Preamble & & PHY Rate & Preamble & PHY Rate & Preamble \\ 
MCS  &  (Mbps)   & ($\mu s$) 
     &  (Mbps)    & ($\mu s$)  & & (Mbps)   & ($\mu s$) & (Mbps) & ($\mu s$) \\
     &  GI$=1.6 \mu s$   &
       & GI$=0.8 \mu s$   &   & & GI$=0.9 \mu s$ & & GI$=0.8 \mu$ &  \\ \hline
     & \multicolumn{2}{c|}{32 stations IEEE 802.11 ax} &  &\multicolumn{6}{c|}{} \\ \hline
 0   &   8.1 & 64.8 &  234.0 & 52.0 &&  6.0 & 20.0 & 48.0 & 20.0 \\ 
 1   &  16.3 & 64.8 &  468.0 & 52.0 && 12.0 & 20.0 & 48.0 & 20.0 \\ 
 2   &  24.4 & 64.8 &  702.0 & 52.0 && 24.0 & 20.0 & 48.0 & 20.0 \\ 
 3   &  32.5 & 64.8 &  936.0 & 52.0 && 24.0 & 20.0 & 48.0 & 20.0 \\ 
 4   &  48.8 & 64.8 & 1404.0 & 52.0 && 48.0 & 20.0 & 48.0 & 20.0 \\ 
 5   &  65.0 & 64.8 & 1872.0 & 52.0 && 48.0 & 20.0 & 48.0 & 20.0 \\ 
 6   &  73.1 & 64.8 & 2106.0 & 52.0 && 48.0 & 20.0 & 48.0 & 20.0 \\ 
 7   &  81.3 & 64.8 & 2340.0 & 52.0 && 48.0 & 20.0 & 48.0 & 20.0 \\ 
8    &  97.5 & 64.8 & 2808.0 & 52.0 && 48.0 & 20.0 & 48.0 & 20.0 \\ 
 9   & 108.3 & 64.8 & 3120.0 & 52.0 && 48.0 & 20.0 & 48.0 & 20.0 \\ 
10   & 121.9 & 64.8 &   N/A  &  N/A && 48.0 & 20.0 &  N/A &  N/A \\ 
11   & 135.4 & 64.8 &   N/A  &  N/A && 48.0 & 20.0 &  N/A &  N/A \\ \hline 
     & \multicolumn{2}{c|}{64 stations IEEE 802.11 ax} &  &\multicolumn{6}{c|}{} \\ \hline
 0   &  3.5 & 64.8 &  234.0 & 52.0 &&  6.0 & 20.0 & 48.0 & 20.0 \\ 
 1   &  7.1 & 64.8 &  468.0 & 52.0 &&  6.0 & 20.0 & 48.0 & 20.0 \\ 
 2   & 10.6 & 64.8 &  702.0 & 52.0 &&  9.0 & 20.0 & 48.0 & 20.0 \\ 
 3   & 14.2 & 64.8 &  936.0 & 52.0 && 12.0 & 20.0 & 48.0 & 20.0 \\ 
 4   & 21.3 & 64.8 & 1404.0 & 52.0 && 18.0 & 20.0 & 48.0 & 20.0 \\ 
 5   & 28.3 & 64.8 & 1872.0 & 52.0 && 24.0 & 20.0 & 48.0 & 20.0 \\ 
 6   & 31.9 & 64.8 & 2106.0 & 52.0 && 24.0 & 20.0 & 48.0 & 20.0 \\ 
 7   & 35.4 & 64.8 & 2340.0 & 52.0 && 24.0 & 20.0 & 48.0 & 20.0 \\ 
 8   & 42.5 & 64.8 & 2808.0 & 52.0 && 36.0 & 20.0 & 48.0 & 20.0 \\ 
 9   & 47.2 & 64.8 & 3120.0 & 52.0 && 36.0 & 20.0 & 48.0 & 20.0 \\ 
10   &  N/A &  N/A &  N/A   &  N/A &&  N/A &  N/A &  N/A &  N/A \\ 
11   &  N/A &  N/A &  N/A   &  N/A &&  N/A &  N/A &  N/A &  N/A \\ \hline 
\end{tabular}  
\end{table}

We assume the
Best Effort Access Category
in which $AIFS=43 \mu s$, $SIFS=16 \mu s$ and $CW_{min}=16$
for the transmissions of both the AP and the stations.
The BackOff interval is a random number
chosen uniformly from
the range $[0,....,CW_{min}-1]$. 
In 11ax  there are no collisions and
since we consider a very 'large' number
of transmissions from the AP, we take the BackOff average
value of  
$\ceil{\frac{CW_{min}-1}{2}}$ and the average BackOff interval is
$\ceil{\frac{CW_{min}-1}{2}} \cdot SlotTime$
which equals $67.5 \mu s$ for a $SlotTime= 9 \mu s$.

Assuming an OFDM based PHY layer every OFDM symbol in IEEE 802.11ac
is $3.2 \mu s$.  We assume also similar multi-path
conditions and therefore set the DL and UL Guard Intervals (GI)
to $0.8 \mu s$. Thus, in IEEE 802.11ac
the duration of a symbol in the DL and UL is $4 \mu s$.
In IEEE 802.11ax the symbol is $12.8 \mu s$. In the DL
we assume again a GI of $0.8 \mu s$ and therefore the symbol
in this direction is $13.6 \mu s$. In the UL MU we assume
a GI of $1.6 \mu s$ and therefore the symbol in this
direction is $14.4 \mu s$. The UL GI is $1.6 \mu s$ 
due to UL arrival time variants. In SU UL the GI is $0.8 \mu s$.

Finally, we assume that the MAC Header field
is of 28 bytes and
the Frame Control Sequence (FCS) field is of 4 bytes.
We also consider several channel conditions which
are expressed by different values of the Bit Error Rate (BER)
which is the probability that a bit arrives corrupted
at the destination. We assume a model where these probabilities
are bitwise independent~\cite{L1}.

%% file: thrana.tex
\clearpage

\section{Throughput analysis}

Let $X$ be the number
of MPDU frames in an A-MPDU frame, numbered $1,..,X$, and $Y_i$ be the number
of MSDUs in MPDU number $i$. 
Let {\it MacHeader, MacDelimiter} and $FCS$ be the length, in bytes,
of the  MAC Header, MAC Delimiter and FCS fields respectively,
and let $O_M = MacHeader + MacDelimiter + FCS$.
Let $L_{DATA}$ be the length, in bytes,
of the MSDU frames.
Also, let 
$Len=4 \cdot \ceil{\frac{L_{DATA}+14}{4}}$ 
and 
$C_i = 8 \cdot 4 \cdot \ceil{\frac{O_M + Y_i \cdot Len}{4}}$.
$C_i$ is the length, in bits, of MPDU number $i$.

\subsection{IEEE 802.11ac}

The throughput of 11ac when only one station
is transmitting in the network, Figure~\ref{fig:traffic}(A),
is given by Eq.~\ref{equ:thrac}~\cite{SA}: 

\begin{equation}
Thr{AC}=
\frac
{\sum_{i=1}^{X} 8 \cdot Y_i \cdot L_{DATA} \cdot (1-BER)^{C_i} }
{AIFS+BO(Variable)+P_{UL}+ T(DATA)+SIFS+P_{DL}+T(BAck)}
\label{equ:thrac}
\end{equation}

where:

\begin{eqnarray}
T(DATA)= TSym_{UL} \cdot \ceil{\frac{\sum_{i=1}^{X} C_i + 22}{TSym_{UL} \cdot R_{UL}}}
\\ \nonumber
T(BAck) = TSym_{DL} \cdot \ceil{\frac{(30 \cdot 8) +22}{TSym_{DL} \cdot R_{DL}}}
\\ \nonumber
\label{equ:timeac}
\end{eqnarray}

$T(DATA)$ and $T(BAck)$ are the transmission times
of the data A-MPDU frames and the
BAck frames respectively.
$T(BAck)$
is based on the BAck frame
format given in Figure~\ref{fig:frameformat}(A) assuming
the acknowledgment of 64 MPDUs per A-MPDU frame.

$TSym_{UL}$ and $TSym_{DL}$ are the lengths
of the OFDM symbols used in the UL and DL respectively
and every transmission
must be of an integral number of OFDM symbols.
The additional 22 bits in the numerators of $T(DATA)$
and $T(BAck)$
are due to the SERVICE and TAIL fields that are added to every
transmission by the PHY layer conv. protocol~\cite{IEEEBase1}.
$R_{DL}$ and $R_{UL}$ are the DL and UL PHY rates respectively
and $P_{DL}$ and $P_{UL}$ are the preambles used in the
DL and UL respectively, see Figure~\ref{fig:formatPPDU}.

Concerning the throughput of 11ac where several
stations transmit over the UL, we use the analysis
in~\cite{CBV} and verify this analysis by simulation.

\subsection{IEEE 802.11ax}

The throughput of 11ax for
the MU case, i.e. 
the traffic pattern in Figure~\ref{fig:traffic}(D),
is given by
Eq.~\ref{equ:thrax}~\cite{SA}:

\tiny

\begin{equation}
Thr{AX}=
\frac
{S \cdot \sum_{i=1}^{X} 8 \cdot Y_i \cdot L_{DATA} \cdot (1-BER)^{C_i} }
{AIFS+BO(Variable)+P_{DL}+ T(TF)+SIFS+P_{UL} +T(DATA)+ PE + SIFS+P_{DL}+T(Mul.BAck)}
\label{equ:thrax}
\end{equation}

\normalsize

where:

\begin{eqnarray}
T(DATA)= TSym_{UL} \cdot \ceil{\frac{\sum_{i=1}^{X} C_i + 22}{TSym_{UL} \cdot R_{UL}}}
\\ \nonumber
T(TF) = TSym_{DL} \cdot \ceil{\frac{((28+(\frac{S}{2} \cdot 5)) \cdot 8) +22}{TSym_{DL} \cdot R_{DL}}}
\\ \nonumber
T(Mul. BAck) = TSym_{DL} \cdot \ceil{\frac{((22+S \cdot 12) \cdot 8) +22}{TSym_{DL} \cdot R_{DL}}}
\\ \nonumber
\label{equ:timeax}
\end{eqnarray}

\normalsize

\indent
$T(DATA)$, $T(TF)$ and $T(Mul.BAck)$ are the transmission times
of the data A-MPDU frames, the TF frame and the Multi Station 
BAck frame respectively.
$T(MUl.BAck)$
is based on the Multi Station BAck frame
length given in Figure~\ref{fig:frameformat}(C) assuming
the acknowledgment of 64 MPDUs per A-MPDU frame.
When considering the acknowledgment of 256 MPDUs
the term 12 in the numerator is replaced by 36.
The term $S$ in $T(TF)$ and $T(Mul. BAck)$ denotes
the number $S$ of stations transmitting data simultaneously
over the UL. 

\indent
Notice that by setting $S=1$ in
the numerator of Eq.~\ref{equ:thrax} and replacing
$T(Mul.BAck)$ 
in the denominator of Eq.~\ref{equ:thrax}
by $T(BAck)$ of Eq.2
we receive
the throughput of the $SU_{AX}(1)$ mode,
Figure~\ref{fig:traffic}(C).
By further deleting the $T(TF)+SIFS$ in the
denominator of Eq.~\ref{equ:thrax} we
receive the throughput of 11ax when only one
station is transmitting in the system $SU_{AX}$,
Figure~\ref{fig:traffic}(A),
the same as Eq.~\ref{equ:thrac}.
Recall that $T(BAck)$ in Eq. 2 assumes
the acknowledgment of 64 MPDUs. In 11ax it is also
possible to acknowledge 256 MPDUs and in this case the 30 bytes
in the numerator of $T(BAck)$ are replaced by 54 bytes, 
see Figure~\ref{fig:frameformat}(B).

$TSym_{UL}$ and $TSym_{DL}$ are the lengths
of the OFDM symbols used in the UL and DL respectively
and every transmission
must be of an integral number of OFDM symbols.
The additional 22 bits in the numerator of $T(DATA)$, $T(TF)$
and $T(Mul.BAck)$
are due to the SERVICE and TAIL fields that are added to every
transmission by the PHY layer conv. protocol~\cite{IEEEBase1}.
$R_{DL}$ and $R_{UL}$ are the DL and UL PHY rates respectively
and $P_{DL}$ and $P_{UL}$ are the preambles used in the
DL and UL respectively (see Figure~\ref{fig:formatPPDU}).
 
The terms in Eqs.~\ref{equ:thrac} and~\ref{equ:thrax} are not continuous and so
it is difficult to find the optimal X and Y, i.e.
the values for X and Y that maximize 
the throughput. However, in~\cite{SA} it is
shown that if one neglects the rounding in the
denominators of Eqs.~\ref{equ:thrac} and~\ref{equ:thrax} then the optimal
solution has the property that all the MPDUs
contain almost the same number of MSDUs: the difference
between the largest and smallest number of MSDUs in MPDUs
is at most 1. The difference is indeed 1
if the limit on the transmission time of the PPDU
does not enable transmission of
the same number of MSDUs in all MPDUs.


We therefore use the result in~\cite{SA} and look for the
maximum throughput as follows: We check for every
X, $1 \le X \le 64$ (also $1 \le X \le 256$ for 11ax)
and for every Y, $1 \le Y \le Y_{max}$, for the
received throughput such that $Y_{max}$ is the
maximum possible number of MSDUs in an MPDU.
All is computed taking into account
the upper limit of $5.484 ms$ on the transmission time
of the PPDU (PSDU+preamble). In case it is not possible
to transmit the same number of MSDUs in all
the MPDUs, some of the MPDUs have one more MSDU
than the others, up to the above upper limit
on transmission time. 

The analytical results of 11ax have been verified by 
an 11ax simulation model running on the $ns3$
simulator~\cite{NS3} and the simulation and analytical
results are the same. This outcome is not surprising however, because
there is not any stochastic process involved
in the scheduled transmissions in 11ax
assumed in this paper. 
Therefore, we do not mention the simulation results any further
in this paper.

%% file: thrres.tex
\section{Throughputs' models and results}

\subsection{Transmissions' models and scenarios}

We compare between all applicable configurations
and scheduling flavors of the stations' 
transmissions up to 64 stations. The scheduling flavors are
as follows.

\noindent
Concerning 11ac :

\begin{itemize}

\item
UL using CSMA/CA . DL Ack transmissions are conducted at the 
basic rate set.

\end{itemize}

\noindent
Concerning 11ax :

\begin{itemize}

\item
UL one station
transmits up to 64 or 256 MPDUs in an A-MPDU frame.
DL Ack and TF transmissions are conducted at the
basic rate set.
When only one station is in the system the AP transmits
the BAck control frame only. When there are several stations
in the system
the AP also transmits the TF control frame.
We denote by
11ax/64 and 11ax/256 the cases when a station
transmits up to 64 or up to 256 MPDUs per 
A-MPDU frame respectively.

\item
UL S= 4, 8, 16, 32 or 64 stations
are transmitting in MU. For $S>4$ also OFDMA is used 
over the UL.
Up to 64 MPDUs or 256 MPDUs are transmitted
in an A-MPDU frame. DL Ack and TF transmissions are conducted
at the basic rate set.

\end{itemize}

For every number $S$ of stations we analyze the optimal
working point, i.e. the one that optimizes the throughput,
as a function of the transmission scheduling flavor,
MCS in use and the A-MPDU frame structure.

First, we checked for every
number of stations all of the possible transmission 
scheduling flavors applicable
for this number of stations. For 11ac only CSMA/CA
is used over the UL  but for 11ax
several transmissions scheduling
flavors are possible. For example, for 64 stations 
one can
use 64 cycles of Figure~\ref{fig:traffic}(C) sequentially
i.e. $64 \cdot SU_{AX}(1)$. One can also use 16 cycles 
of Figure~\ref{fig:traffic}(D),
namely $16 \cdot MU_{AX}(4)$.
Finally, one can also use 8, 4, 2 and 1 
cycles of Figure~\ref{fig:traffic}(D)
denoted before $MU_{AX}(8)$, $MU_{AX}(16)$, $MU_{AX}(32)$ 
and $MU_{AX}(64)$
respectively. 

Every transmission scheduling flavor is checked over all the
applicable MCSs. For 11ac these are MCS0-MCS9. For 11ax
these are MCS0-MCS11 except in the
case of 64 stations where only MCS0-MCS9 are
applicable. We also check for every transmission scheduling flavor
and MCS the optimal working point by optimizing
the number of MPDUs in A-MPDU frames
and the number of MSDUs
in every MPDU that yields the maximum throughput, i.e. we look
for the optimal A-MPDU frame structure.

We checked all the above for MSDUs of 64, 512 and 1500 bytes
and BER$=$$0, 10^{-5}$.

In the next section we show three sets of results.
In Figure~\ref{fig:res1} we show the maximum throughputs
received for every number of stations in every transmission
scheduling flavor for MSDUs of 1500 bytes. The results for
MSDUs of 64 and 512 bytes are similar. In Figure~\ref{fig:res2}
we demonstrate for $MU_{AX}(4)$ and $MU_{AX}(64)$
the maximum
throughputs received in the
various MCSs
and the influence of the maximum number of MPDUs per
A-MPDU frame, 64 or 256, on the received throughput,
both for BER$=$0 and BER$=$$10^{-5}$.
Finally, in Figure~\ref{fig:res3} we show
the influence of the number of MPDUs in an A-MPDU frame,
from 1 to 256, on the
received throughput for the case of $MU_{AX}(4)$ and 
$MU_{AX}(64)$, both
for BER$=$0 and BER$=$$10^{-5}$.

\subsection{Throughput results}

\begin{figure}
\vskip 16cm
\includegraphics{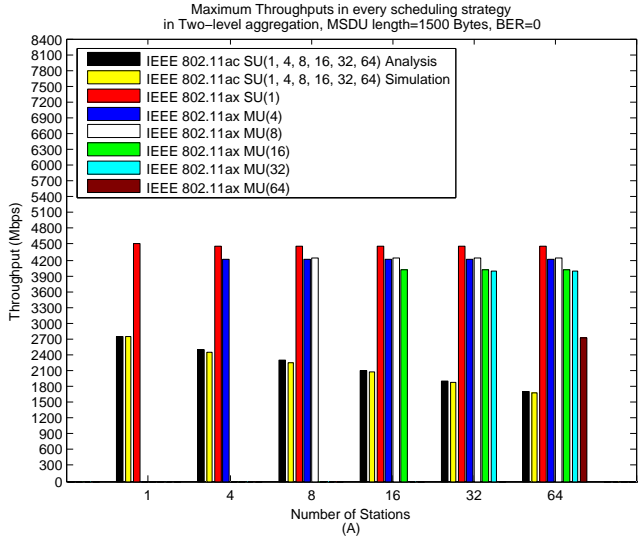}
\includegraphics{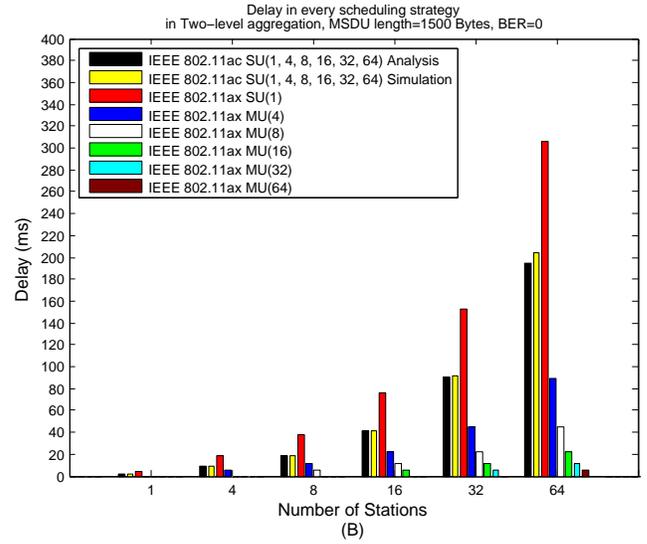}
\includegraphics{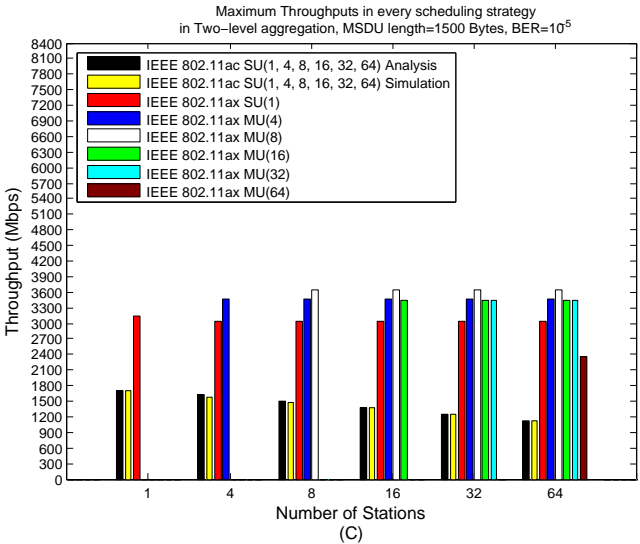}
\includegraphics{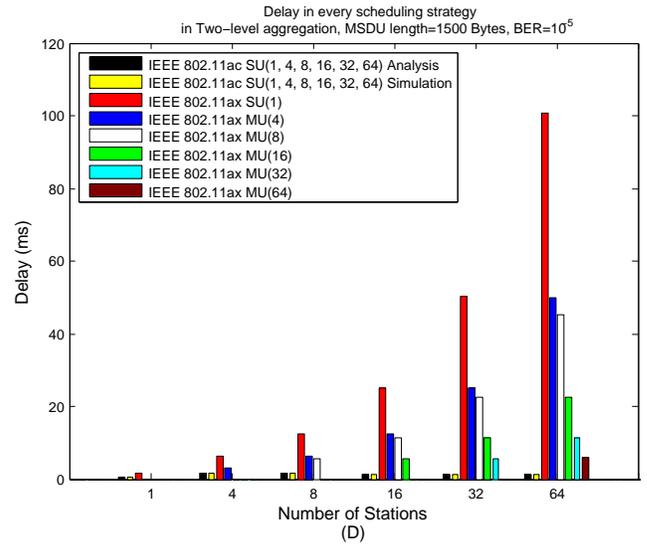}
\caption{Maximum throughputs and corresponding delays in Single User and Multi User Uplink transmissions in IEEE 802.11ac and IEEE 802.11ax .}
\label{fig:res1}
\end{figure}

Recall that in Figure~\ref{fig:res1} we show the maximum throughputs
received as a function of the number
of transmitting stations. 
We show results for MSDUs of 1500 bytes only;
similar results are received for MSDUs
of 64 and 512 bytes. 

For 11ac we show analytical results received 
from the analysis in~\cite{CBV} and we also verify
these results by simulation. For
11ax, when there is only one station in the network
we use Figure~\ref{fig:traffic}(A) to compute
the received throughput. When several stations
transmit one at a time, we use the transmission pattern in
Figure~\ref{fig:traffic}(C) to obtain the
results. However, in the legend of all the graphs
in Figure~\ref{fig:res1}, these results are shown
together under 11ax SU(1).

In Figure~\ref{fig:res1}(A) we show results for
BER$=$0. When referring in the legend to e.g. 11ax MU(4) 
we refer to $MU_{AX}(4)$, i.e. the case in which 4 stations
transmit
simultaneously to the AP
using UL MU, Figure~\ref{fig:traffic}(D).
When showing results for $MU_{AX}(4)$ in
the case of e.g. 64 stations, the traffic
cycle in Figure~\ref{fig:traffic}(D) repeats
itself 16 times. In every cycle a different group
of 4 stations is transmitting, i.e. $16 \cdot MU_{AX}(4)$.

Note that for 11ac the analytical and simulation
results match very closely. For one station in the
network, the traffic pattern in Figure~\ref{fig:traffic}(A),
11ax has a much larger throughoput than 11ac
because in 11ax it is possible to transmit A-MPDU frames of
256 MPDUs, while in 11ac the number of MPDUs per A-MPDU frame is
limited to 64.  11ax outperforms 11ac by 64$\%$.

We see in Figure~\ref{fig:res1}(A) that the largest
throughput is received in $SU_{AX}(1)$.
Notice however that the throughput of $SU_{AX}(1)$ when
only one station is transmitting in the system
is larger than the throughput of $SU_{AX}(1)$ 
when $S>1$ stations are transmitting.
This is due to
the lack of the TF frame when one
station transmits, and using
the BAck frame which is shorter than the Multi Station BAck.
$SU_{AX}(1)$ has the largest throughput among
all transmission scheduling flavors because of its
relatively larger PHY rate - it is larger than 4 times the one in
the case of 4 stations, larger than 8 times the one in the
case of 8 stations etc.

The throughout of $MU_{AX}(8)$ is the same as that of
$MU_{AX}(4)$.
From Table 2 one can see that
the PHY rates in $MU_{AX}(8)$ are half of those in $MU_{AX}(4)$. This 
is balanced by twice the number of stations that are transmitting.

The throughput of $MU_{AX}(16)$ is smaller than that
of $MU_{AX}(8)$ because its PHY rates are less
than half
those of $MU_{AX}(8)$.
The throughput of $MU_{AX}(32)$ is less than that
in $MU_{AX}(16)$ although its PHY rates are half
those in $MU_{AX}(32)$ due to the transmission time
of the TF frame. In the case of 16 stations it is one 
symbol while in 32 stations it is two symbols.

The throughput of $MU_{AX}(64)$ is the smallest 
because of its very small PHY rates which are
much less than half those in $MU_{AX}(32)$.
Recall also that MCS10 and MCS11 are not applicable in the
case of 64 stations. Also, the transmission of the TF
frame now requires 7 symbols.

Finally, 11ax outperforms 11ac by $78 \%$ and
$263 \%$ for 4 and 64 stations respectively
because 11ax uses a scheduled transmission pattern
while 11ac is based on the contention
CSMA/CA MAC protocol access with collisions.

Although the throughput metric is important, the access
delay metric is also important. This
metric is defined in this paper as the
time elapsed between two consecutive transmissions from
the same station to the AP. 

In Figure~\ref{fig:res1}(B) we show the access delay
for the various transmissions' scheduling flavors. 
Some applications benefit primarily from lower
latency, especially real-time streaming applications
such as voice, video conferencing or even video chat.
The trade-offs between latency and throughput becomes
more complex as applications are scaled out to run 
in a distributed fashion. 
The access delay results are
as expected; the access delay is lower when
more stations transmit simultaneously. It seems that except
for $SU_{AX}(1)$ the
cycles are about the same length in all the
transmissions' scheduling flavors and the relation between
the access delays is about the same relation between
the number of stations transmitting simultaneously.
An exception is the case of $SU_{AX}(1)$ where a cycle is shorter, $4.7 ms$
vs. $5.6 ms$ in the other cases and therefore in the case
of e.g. 64 stations, the relation between the access delay
of $SU_{AX}(1)$ and $MU_{AX}(64)$ is about 53.

In Figure~\ref{fig:res1}(C) we show the maximum throughput
as a function of the number of stations for the case
BER$=$$10^{-5}$. An interesting difference compared to
BER$=$0 is that the throughput of $MU_{AX}(4)$ is larger
than that of $SU_{AX}(1)$. The reason for this phenomena
is the relation
between the PHY rates in both
schemes to the overhead. In BER$=$0 the large PHY rate in 
$SU_{AX}(1)$ causes the 
overhead as the AIFS and BO to be significant and this
leads to large MPDUs in order to achieve a large
throughput. On the other hand in BER$=$$10^{-5}$
it is efficient to transmit short MPDUs in order
to achieve a large MPDU transmission success probability
and therefore to
a larger throughput. 
It turns out that in $SU_{AX}(1)$
it is most efficient to transmit
MPDUs of two MSDUs but in $MU_{AX}(4)$ MPDUs of one MSDU
are the most efficient. Overall the larger 
MPDUs' success probability in $MU_{X}(4)$ together with
the smaller PHY rate makes $MU_{AX}(4)$ more efficient.

Notice also that $MU_{AX}(8)$ outperforms
$MU_{AX}(4)$. This occurs due to its short MPDUs and the smaller
PHY rates. The optimal A-MPDU frame structure in $MU_{AX}(4)$
is 256 MPDUs of one MSDU each while in $MU_{AX}(8)$ it is
242 MPDUs of one MSDU each.
In $MU_{AX}(4)$ a cycle
lasts $3.11 ms$ and in $MU_{AX}(8)$ it lasts $5.63 ms$. 
In $MU_{AX}(8)$ almost twice the number of MSDUs are transmitted
than in $MU_{AX}(4)$, but this is done in shorter than twice
the cycle length of $MU_{AX}(4)$.
This leads to a larger throughput in $MU_{AX}(8)$.
The reasons why $MU_{AX}(32)$ and $MU_{AX}(64)$ have
the smallest throughputs are explained earlier as in the
case of BER$=$0.

In the case of a single transmitting station 11ax
outperforms 11ac by 85$\%$. $MU_{AX}(8)$ outperforms
11ac by 270$\%$.

In Figure~\ref{fig:res1}(D) we show the corresponding access
delays for the transmissions' scheduling flavors for BER$=$$10^{-5}$.
Worth mentioning is the relation between the access delays
of $MU_{AX}(4)$ and $MU_{AX}(8)$. 
For BER$=$$10^{-5}$ they are close to each other
because the maximum throughput in both scheduling
flavors is received when A-MPDU frames contain 255 and 242 MPDUs
respectively of 1 MSDU each. Since the PHY rates in $MU_{AX}(8)$ are 
half those in $MU_{AX}(4)$, the cycle length in $MU_{AX}(8)$ is
about double in length as in $MU_{AX}(4)$. However, this
is compensated by double the number of stations to which
the AP transmits in $MU_{AX}(8)$ compared to $MU_{AX}(4)$; the
overall is similar access delays in both scheduling flavors.
This situation is different than that of BER$=$0.
In BER$=0$ the cycle length in both $MU_{AX}(4)$ and $MU_{AX}(8)$ is
about the same, around $5.5 ms$, transmitting as many MSDUs
as possible. The limiting factor on the
cycle length is the limit on the
transmission time of a PPDU.
The access delay in $MU_{AX}(4)$ is now twice that
of $MU_{AX}(8)$ because of the 4 vs. 8 transmitting stations
in $MU_{AX}(4)$ and $MU_{AX}(8)$ respectively.

The access delays for BER$=$$10^{-5}$ are smaller
than those of BER$=$0 because the MPDUs are shorter, usually containing
one MSDU compared to 7 MSDUs in BER$=$0.
However, the throughputs are also lower.

In overall $MU_{AX}(16)$ and $MU_{AX}(32)$ seem to be
the best transmission scheduling flavors achieving large
throughputs with small access delays.

\begin{figure}
\vskip 16cm
\includegraphics{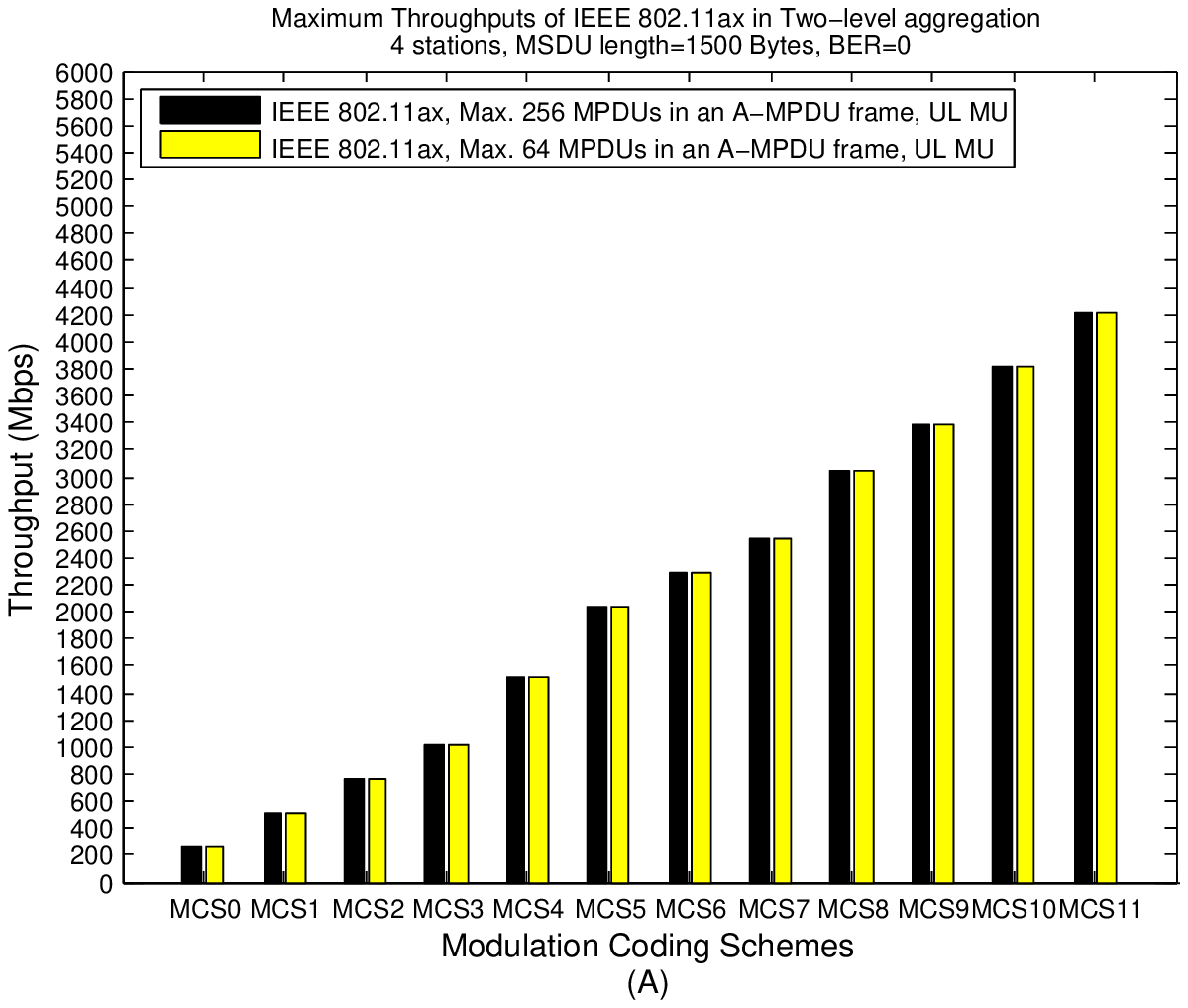}
\includegraphics{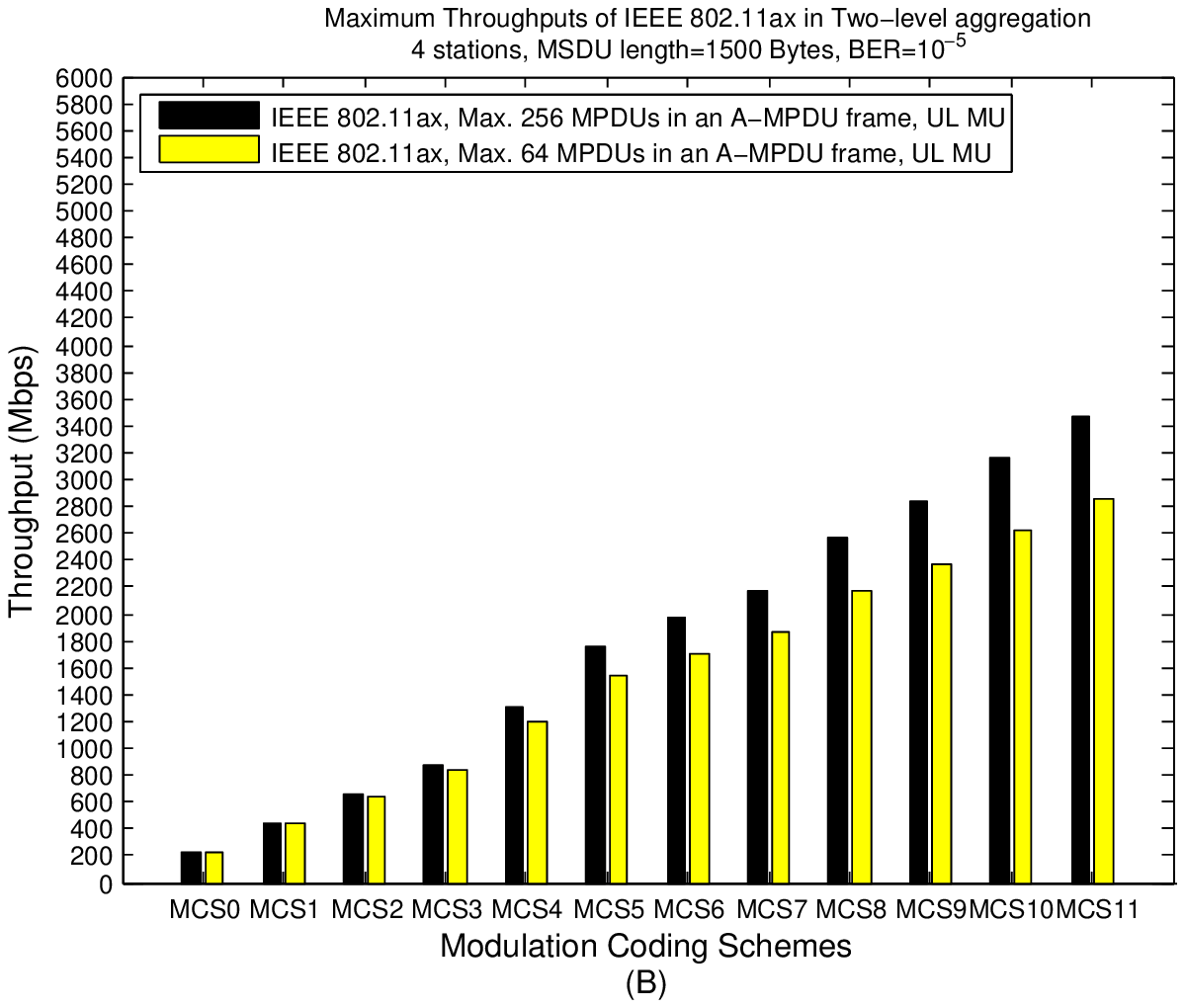}
\includegraphics{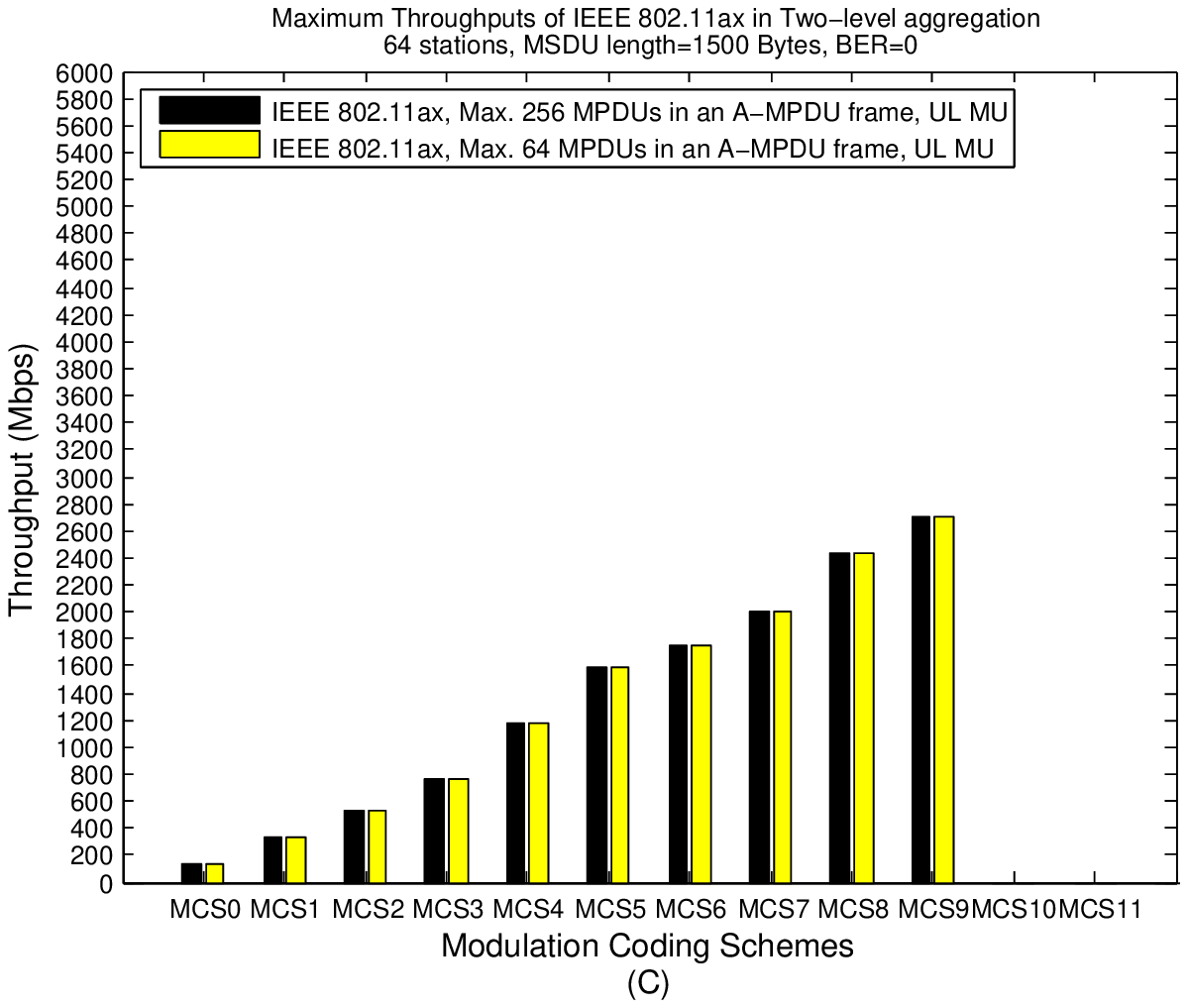}
\includegraphics{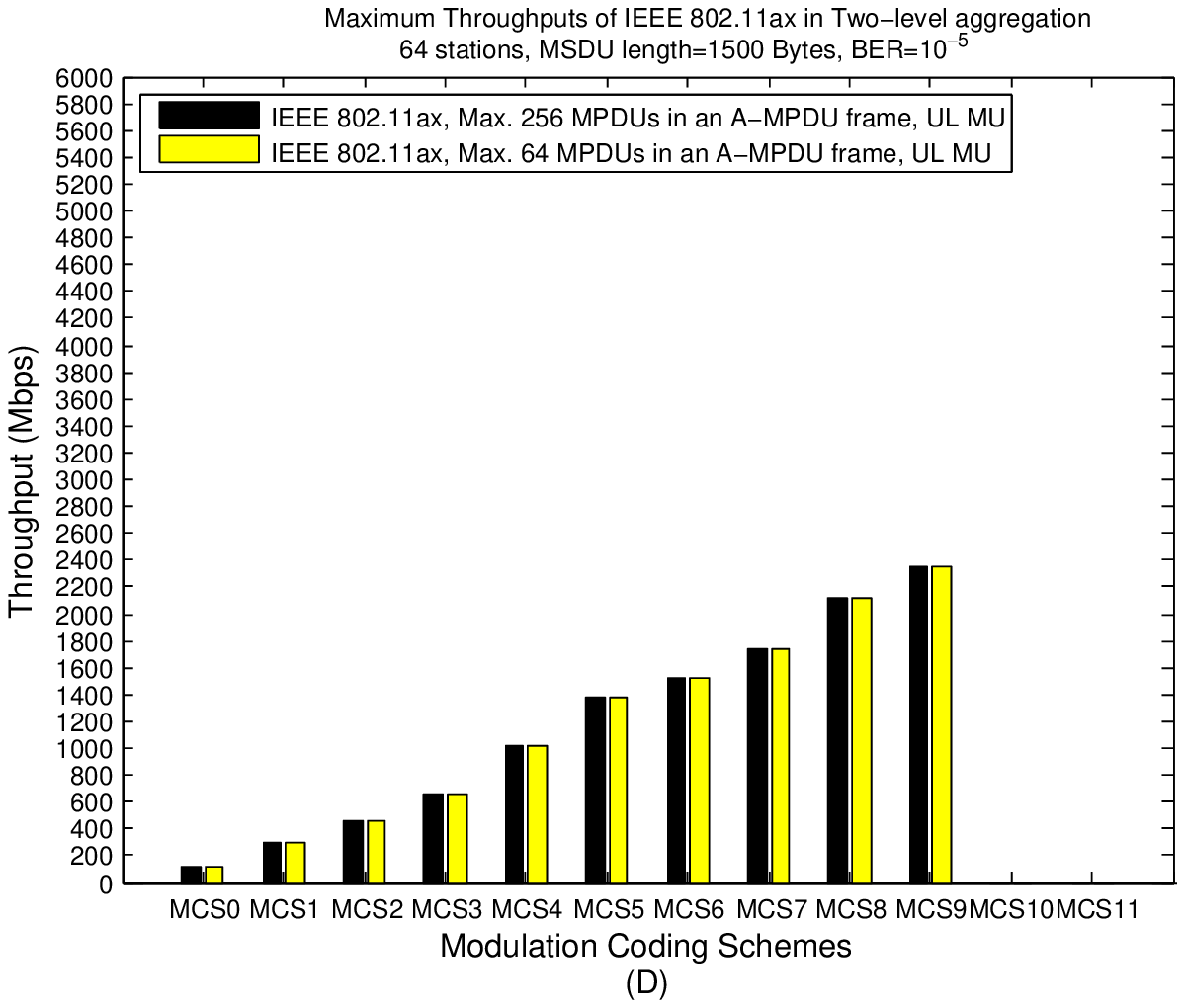}
\caption{The throughputs in IEEE 802.11ax when 4 and 64 stations transmit simultaneously to the AP, as a function of the MCSs and the number of MPDUs in A-MPDU frames. }
\label{fig:res2}
\end{figure}

In Figure~\ref{fig:res2} we show the throughput performance
of $MU_{AX}(4)$ and $MU_{AX}(64)$ for every MCS, for
BER$=$0 and $10^{-5}$, and for the cases using 64 and 256 MPDUs
per A-MPDU frame. In Figures~\ref{fig:res2}(A) 
and~\ref{fig:res2}(B) we show
the results for $MU_{AX}(4)$ for BER$=$0 and BER$=$$10^{-5}$ respectively.
In Figures~\ref{fig:res2}(C) 
and~\ref{fig:res2} (D) the same results
respectively are shown for $MU_{AX}(64)$.
Notice that for $MU_{AX}(64)$ there are no results for MCS10 and MCS11
which are not applicable in this case due to the small
PHY rates.

The maximum throughput in $MU_{AX}(4)$
is always received
in MCS11 ( MCS9 in $MU_{AX}(64)$ ) due to the
largest PHY rates in this MCS. Considering $MU_{AX}(4)$ notice
that for BER$=$0 11ax/256 outperforms 11ax/64 only in MCS10 and MCS11
while in BER$=$$10^{-5}$ 11ax/256 outperforms 11ax/64 starting
from MCS2.
In BER$=$0 it is efficient to transmit large MPDUs. Therefore,
the limit on the A-MPDU frame size is imposed by the limit
of $5.484 ms$ on the transmission time of the PPDU. Only
in larger PHY rates there is room for more than  64 MPDUs and
in these cases 11ax/256 has an advantage over 11ax/64 . In BER$=$$10^{-5}$
it is efficient to transmit short MPDUs. In this case the significant
limit is the number of MPDUs. 11ax/256 outperforms 11ax/64
from MCS2 because it enables transmitting more short MPDUs than
11ax/64 .

In $MU_{AX}(64)$ there is no difference between
11ax/256 and 11ax/64 because the small PHY rates do not enable
transmission of more than 64 MPDUs in every
MCS, given the limit
of the $5.484 ms$ on the transmission time of the PPDU.

\begin{figure}
\vskip 16cm
\includegraphics{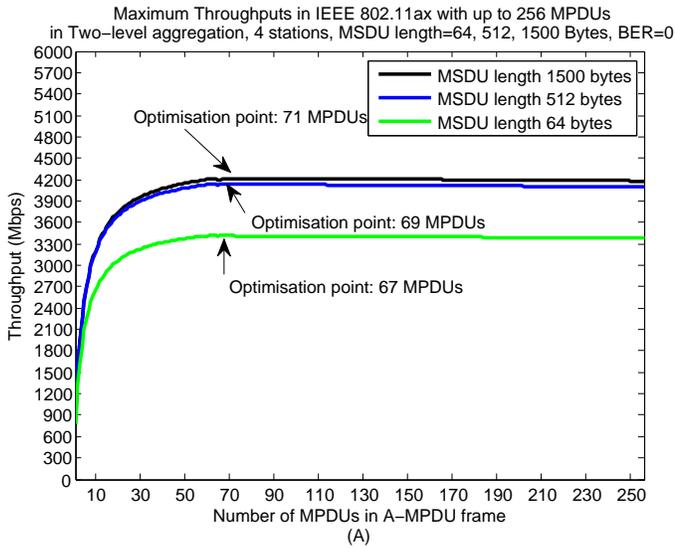}
\includegraphics{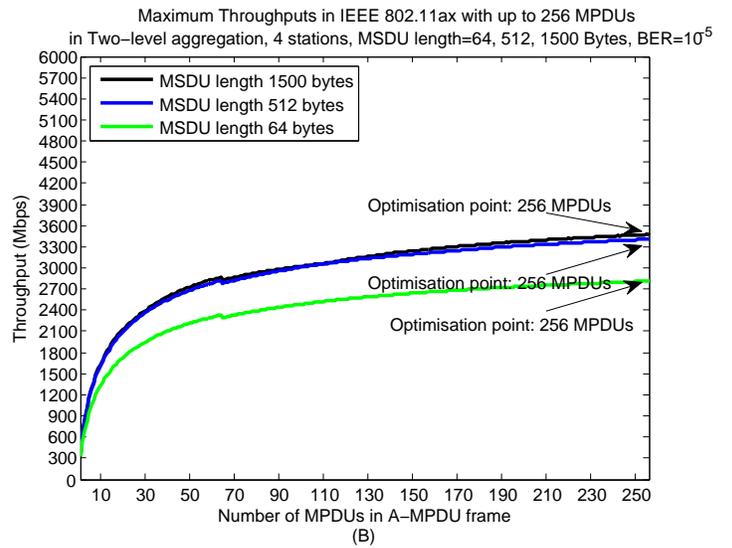}
\includegraphics{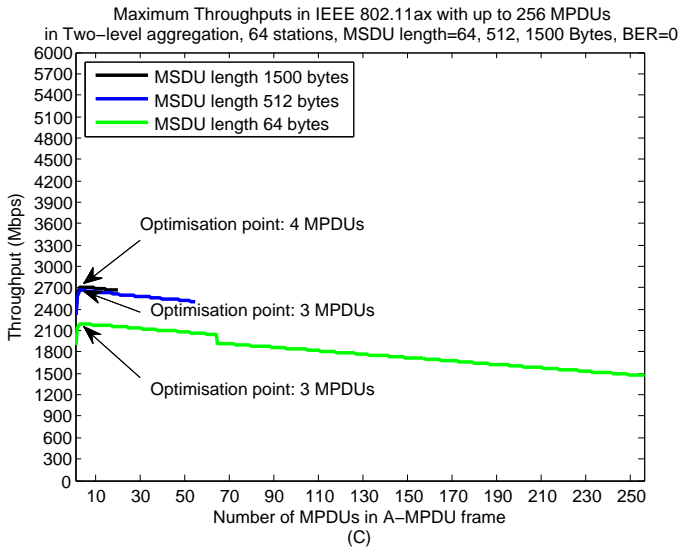}
\includegraphics{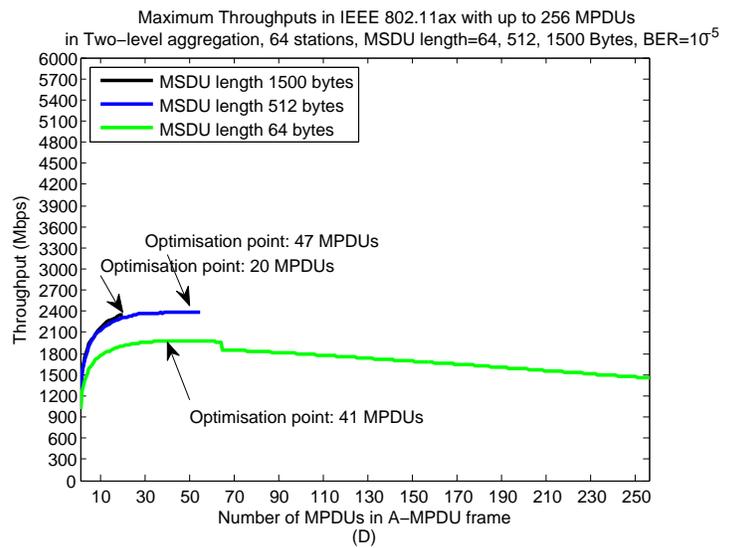}
\caption{The throughputs vs. the number of MPDUs in A-MPDU frames in IEEE 802.11ax Multi User for 4 stations in MCS11 and 64 stations in MCS9.}
\label{fig:res3}
\end{figure}

In Figure~\ref{fig:res3} we show the impact of the number
of MPDUs in A-MPDU frames on the received throughput.
In Figures~\ref{fig:res3}(A) and~\ref{fig:res3}(B) results are
shown for $MU_{AX}(4)$ in MCS11, for BER$=$0 and BER$=$$10^{-5}$
respectively. Similar results are shown for $MU_{AX}(64)$ for MCS9
in Figures~\ref{fig:res3}(C) and~\ref{fig:res3}(D) 
respectively. We show results
for MSDUs of 64, 512 and 1500 bytes.

Considering $MU_{AX}(4)$ and BER$=$0, Figure~\ref{fig:res3}(A), there
is an optimal number of MPDUs of around 70 for all the
sizes of the MSDUs. In BER$=$0 it
is efficient to transmit the largest MPDUs as possible.
For about 70 MPDUs all the MPDUs contain the
largest possible number of MSDUs and the transmission
time is used efficiently. Above 70 MPDUs the limit
of $5.484 ms$ on the PPDU transmission time
and the MPDUs' overhead cause a smaller number
of MSDUs to be transmitted and the throughput decreases.

In the case of BER$=$$10^{-5}$, Figure~\ref{fig:res3}(B), the
optimal number of MPDUs is 256 since MPDUs are short 
(to increase the MPDUs' transmission success probability)
and there is
enough transmission time for 256 MPDUs in the A-MPDU frame.
Every additional MPDU increases the throughput.

In $MU_{AX}(64)$, Figures~\ref{fig:res3}(C) 
and~\ref{fig:res3}(D), the PHY rates are smaller
and the limit on the PPDU transmission time does
not enable transmission of many MPDUs with MSDUs of 512 and 1500 bytes.
Up to 20 and 55 MPDUs of these sizes can be transmitted 
respectively, containing one MSDU. For BER$=$0 there is an
optimal number of around 3-4 MPDUs that yields the
maximum throughput for all MSDUs' sizes. A larger number
of MPDUs decreases the number of MSDUs transmitted
due the the MPDUs' overhead and the throughput decreases.
In the case of BER$=$$10^{-5}$ the MPDUs are shorter, 
and increasing the number of MPDUs increases the throughput
since more MSDUs are transmitted.
An exception is the case of 64 bytes MSDUs. In this
case it is possible to transmit 256 MPDUs and several
MSDUs can be transmitted in every MPDU. Increasing
the number of MPDUs in this case decreases the number
of MSDUs transmitted with a decrease in the throughput.

%% file: summary.tex
\newpage

\section{Summary}

In this paper we explore multiple scheduling
strategies in order to compare between the throughputs
of 11ac and 11ax over the Uplink when considering 
UDP traffic and that several stations are transmitting
in the system. 

IEEE 802.11ax outperforms 11ac by the order of several tenths
of percent mainly due to its scheduling strategies
vs.
the SU air access based on
the CSMA/CA contention method in 11ac .
In 11ax the best transmission scheduling flavors are $MU_{AX}(4)$
and $MU_{AX}(8)$ achieving good results in terms of both
throughput and access delay. IEEE 802.11ax achieves its best
throughputs in the largest MCS possible, MCS11 for
up to 32 stations and MCS9 for 64 stations.

There is an optimal working point
for every scheduling strategy in terms of the
A-MPDU frame structure. In $MU_{AX}(4)$
it is sufficient to transmit around 70 and 256 MPDUs
per A-MPDU frame for BER$=$0 and BER$=$$10^{-5}$ respectively.
For $MU_{AX}(64)$ these numbers of MPDUs are smaller, around 10 
and 40 respectively, due to the smaller PHY rates.

Finally, using up to 256 MPDUs in an A-MPDU frame outperforms
the use of up to 64 MPDUs in cases when the PHY rates
are larger and/or the channel is unreliable, i.e. BER$=$$10^{-5}$.

%% file: main1.bbl
\begin{thebibliography}{10}

\bibitem{IEEEBase1}
\newblock{IEEE Std. 802.11$^{TM}$-2016},
\newblock{IEEE Standard for Information Technology - 
Telecommunications and Information Exchange between Systems - Local
and Metropolitan Area Networks - Specific Requirements. Part 11:
Wireless LAN Medium Access Control (MAC) and Physical Layer (PHY)
Specifications},
\newblock{IEEE, NewYork, December 2016.}


\bibitem{IEEEax}
\newblock{IEEE P802.11ax$^{TM}$/D1.2},
\newblock{IEEE Draft Standard for Information Technology - 
Telecommunications and Information Exchange between Systems - Local
and Metropolitan Area Networks - Specific Requirements. Part 11:
Wireless LAN Medium Access Control (MAC) and Physical Layer (PHY)
Specific requirements. }
\newblock{IEEE, NewYork, 2016.}

\bibitem{IEEEac}
\newblock{IEEE Std. 802.11ac$^{TM}$-2013},
\newblock{IEEE Standard for Information Technology - 
Telecommunications and Information Exchange between Systems - Local
and Metropolitan Area Networks - Specific Requirements. Part 11:
Wireless LAN Medium Access Control (MAC) and Physical Layer (PHY)
Specific requirements. Amendment 4: Enhancements for Very
High Throughput for Operation in Bands below 6 GHz},
\newblock{IEEE, NewYork, 2013)}.

\bibitem{PS}
Perahia, E. and Stacey, R. (2013)
\newblock{Next Generation Wireless LANs: 802.11n and 802.11ac .}
\newblock{2nd Edition, Cambridge Press, Cambridge.}

\bibitem{KKL}
Khorov, E., Kiryanov, A. and Lyakhov, A. (2015)
\newblock{IEEE 802.11ax: How to Build High Efficiency WLANs,}
\newblock{\it 2015 International Conference on Engineering
and Telecommunication (Ent).}
\newblock{Moscow, 18-19 November 2015 14-19.}

\bibitem{AVA}
Afaqui, M. S., Villegas, E. G. and Aguilera, E. L. (2016)
\newblock{IEEE 802.11ax: Challenges and Requirements for Future
High Efficiency WiFi.}
\newblock{\it IEEE Wireless Communications 99,  2-9.}

\bibitem{DCC}
Deng, D. J., Chen, K. C. and  Cheng, R. S. (2014)
\newblock{IEEE 802.11ax: Next Generation Wireless Local Area Networks.}
\newblock{\it 2014 10th International Conference
on Heterogeneous Networking for Quality, Security and Robustness (QSHINE),}
\newblock{Rhodes, 18-20 August 2014, 77-82. 
https://doi.org/10.1109/qshine.2014.6928663 .}

\bibitem{B}
Bellalta, B. (2016)
\newblock{IEEE 802.11ax: High-efficiency WLANs.}
\newblock{\it IEEE Wireless Communications, 23, 38-46.
https://doi.org/10.1109/MWC.2016.7422404 .}

\bibitem{KCC}
 Karmakar, R., Chattopadhyay, S. and Chakraborty, S. (2017)
\newblock{Impact of IEEE 802.11n/ac PHY/MAC High Throughput
Enhancement over Transport/Application layer protocols - A Survey.}
\newblock{\it IEEE Communication surveys and tutorials.}


\bibitem{QLYY}
Qu, Q., Li, B., Yang, M. and  Yan, Z. (2015) 
\newblock{An OFDMA based Concurrent Multiuser MAC for Upcoming IEEE 802.11ax.}
\newblock{\it IEEE Wireless Communication and Networking Conference
Workshops (WCNCW) 2015, 136-141.}


\bibitem{LLYQYZY}
Lin, W., Li, B., Yang, M., Qn, Q., Yan, Z., Zuo, X., and Yang, B. (2016)
\newblock{Integrated Link-System level Simulation Platform for the
Next Generation WLAN - IEEE 802.11ax.}
\newblock{\it 2016 IEEE Global Communications Conference (Globecom), 1-7.}


\bibitem{LDC}
Lee, J.,  Deng, D. J. and Chen, K. C. 
\newblock{OFDMA-based hybrid channel access for IEEE 802.11ax WLAN.}
\newblock{\it unpublished.}

\bibitem{KBPSL}
Karaca, M., Bastani, S., Priyanto, B. E., Safavi, M. and Landfeldt, B. (2016)
\newblock{Resource Management for OFDMA based Next Generation 802.11ax
WLANs.}
\newblock{\it 2016 9th IFIP Wireless and Mobile Networking Conference (WMNC).}


\bibitem{JS}
Jones, V. and Sampath, H. (2015)
\newblock{Emerging technologies for WLAN.}
\newblock{\it IEEE Communication Magazine, 5,  141-9.}

\bibitem{RFBBO}
Sanabria-Russo, L., Faridi, A., Bellalta,B., Barcelo, J. and  Oliver, M. (2013)
\newblock{Future evolution of CSMA protocols for the IEEE 802.11
standard.}
\newblock{\it 2013 IEEE International Conference on Communication (ICC), 1274-9 .}

\bibitem{RBFB}
Sanabria-Russo, L., Barcelo, J., Faridi, A. and Bellalta, B. (2014)
\newblock{WLANs throughput improvement with CSMA/ECA .}
\newblock{\it 2014 IEEE Conference on Computer Communication 
Workshops (INFOCOM WKSHPS), 125-6.}

\bibitem{HYSG}
He, Y., Yuan, R.,  Sun, J. and  Gong, W. (2009)
\newblock{Semi-Random Backoff: Towards resource reservation for
channel access in wireless LANs.}
\newblock{\it 2009 IEEE International Conference
 on Network Protocols (ICNP), 21-30.}

\bibitem{KLL}
Khorov, E., Loginov, V. and  Lyakhov, A. (2016)
\newblock{Several EDCA Parameters Sets for Improving Channel Access in
IEEE 802.11ax Networks.}
\newblock{\it 2016 International Symposium on Wireless Communication Systems (ISWCS), 419-423.}



\bibitem{SA}
Sharon, O. and  Alpert, Y. (2014)
\newblock{MAC level Throughput comparison: 802.11ac vs. 802.11n .}
\newblock{\it Physical Communication, 12, 33-49 .}

\bibitem{L1}
Lemmon, J. (2002)
\newblock{Wireless link statistical bit error rate model.}, 
\newblock{Technical Report 02-934, U.S. Department of Commerce,
June, 2002.}

\bibitem{CBV}
Chatzimisios, P., Boucouvalas, A. C. and  Vitas, V. (2003)
\newblock{IEEE 802.11 Packet Delay - A Finite Retry Limit Analysis.}
\newblock{\it 2003 IEEE Global Communications Conference (Globecom), 950-954.}

\bibitem{NS3}
\newblock{ns3 simulator, https://www.nsnam.org}

\end{thebibliography}
